\documentclass[graybox]{svmult}
\usepackage{amsfonts}
\usepackage{amsmath}

\usepackage{url}
\usepackage{float}
\usepackage{amsmath}
\usepackage{amssymb}
\usepackage{tikz}
\usepackage{helvet}         
\usepackage{courier}        
\usepackage{type1cm}        
\usepackage{makeidx}         
\usepackage{graphicx}        
\usepackage{multicol}        
\usepackage[bottom]{footmisc}

\usepackage{comment}
\usepackage{hyperref}

\makeindex             

\begin{document}

\title*{Flavored aspects of QCD thermodynamics from Lattice QCD}
\titlerunning{Flavored aspects of QCD thermodynamics from Lattice QCD}
\author{Olaf Kaczmarek}
\institute{Fakult\"at f\"ur Physik, Universit\"at Bielefeld, D-33615 Bielefeld, Germany\\
\email{okacz@physik.uni-bielefeld.de}}

\maketitle

\abstract{We discuss recent progress in lattice QCD studies on various aspects involving strange quarks.
Appropriate combinations of conserved net strange and net charm fluctuations and their correlations with other conserved charges provide evidence that in the hadronic phase so far unobserved hadrons contribute to the thermodynamics and need to be included in hadron resonance gas models. 
In the strange sector this leads to significant reductions of the chemical freeze-out temperature of strange hadrons. 
In this context, a discussion of data from heavy-ion
collisions at SPS, RHIC and LHC on the chemical freeze-out of hadronic species is
presented. 
It can be observed that a description of the thermodynamics of open strange and open charm degrees of freedom in terms of an uncorrelated hadron gas is valid only up to temperatures close to the chiral crossover temperature. 
This suggests that in addition to light and strange hadrons also open charm hadrons start to dissolve already close to the chiral crossover. 
Further indications that open charm mesons start to melt in the vicinity of $T_c$ is obtained from an analysis of screening masses, while in the charmonium sector these screening masses show a behavior compatible with a sequential melting pattern.
At the end of this chapter we will discuss some basics of lattice gauge theory and Monte Carlo calculations. This will provide the required knowledge for performing first lattice calculations for SU(3) pure gauge theory and studying thermodynamic quantities in the exercises of this chapter.  
Recent progress in extracting spectral and transport properties from lattice QCD calculations will be addressed 
separately in the following chapter.
}


\section{Introduction}

In these lecture notes we will discuss recent lattice QCD\index{QCD!lattice} results relevant for the understanding of strongly interacting matter under extreme conditions. We will mainly focus here on observables which are relevant for the study of such matter at high temperatures and densities that are probed in heavy ion collision experiments.
We will only give a brief introduction to lattice QCD in the next section and refer to the textbooks 
\cite{Gattringer:2010zz,Montvay:1994cy,Rothe:1992nt,DeGrand:2006zz}
and lecture notes \cite{Karsch:2003jg}
for more detailed introductions to lattice field theory. For the topics addressed in this lecture note we also like to refer to the overview articles
on QCD thermodynamics and the QCD phase transition\index{QCD!phase transition}
\cite{Karsch:2003jg, Ding:2015ona, Guenther:2020jwe}
and quarkonium in extreme conditions \cite{Rothkopf:2019ipj}.

\section{QCD Thermodynamics and strangeness}

\subsection{$T_c$ and the equation of state}

At vanishing chemical potentials it is well established that the transition from hadronic matter at small temperatures to the deconfined and chiral symmetric phase at high temperatures is not a real phase transition but rather a smooth cross-over for physical values of the quark masses \cite{Bazavov:2011nk,Bhattacharya:2014ara,Aoki:2006we}.
Strictly speaking an order parameter for the chiral symmetry restoration\index{chiral symmetry!restoration} and a critical temperature\index{critical!temperature} of the chiral phase transition can only be defined in the limit of vanishing light quarks masses where the QCD chiral transition is likely to become a second order phase transition. In the vicinity of the second order phase transition the behavior of chiral observables can be described by scaling properties of the corresonding universality class, which for QCD in the chiral limit is most likely to be the 3D $\mathrm O(4)$ universality class\index{universality class} \cite{HotQCD:2019xnw}. Although physical quark masses may lie outside of this scaling region, nevertheless, they show signatures of pseudo-critical behavior and can be used to define a pseudo-critical temperature, $T_c$, for the chiral cross-over\index{chiral!crossover}.  

\begin{figure}[htb]
\centering{
 \includegraphics[width=0.48\textwidth]{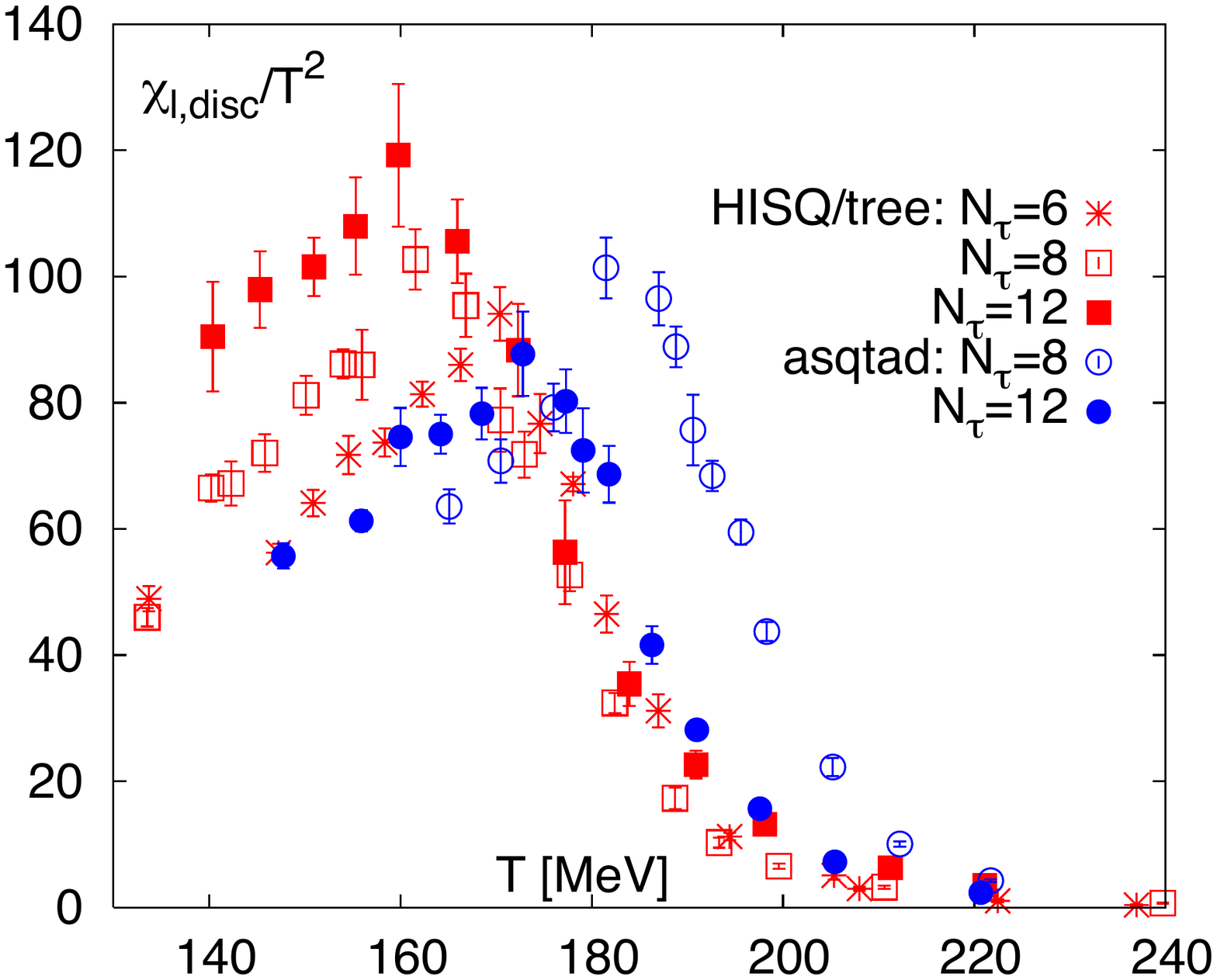}
 \includegraphics[width=0.48\textwidth]{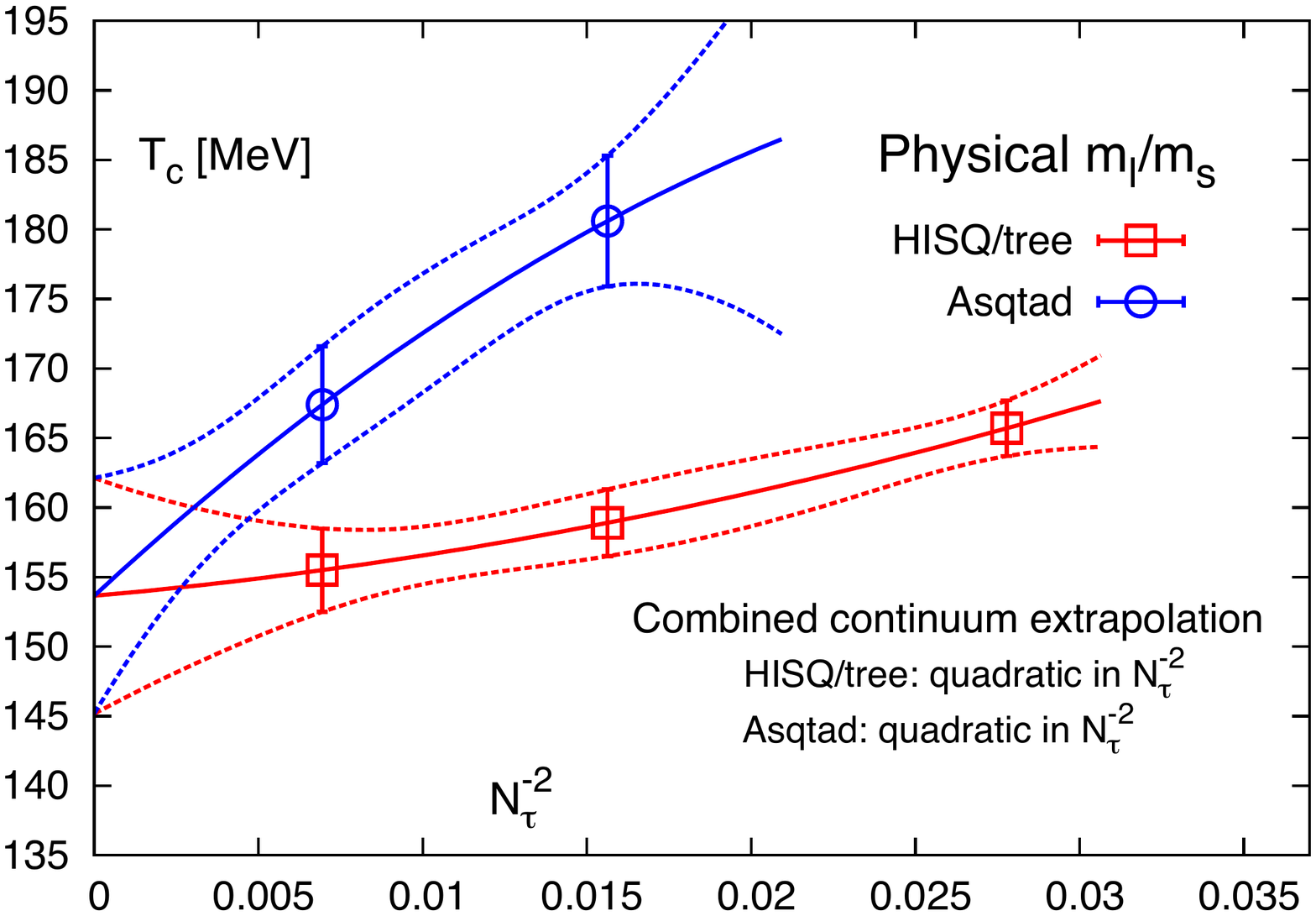}}
\caption{Lattice determination of pseudo-critical temperature $T_c$. 
Left panel: Chiral susceptibility for light flavors $\chi_l$ as a function of temperature for QCD with 2+1 flavors and physical quark masses. 
Right panel: Dependence of $T_c$ on the number of Lattice sites in the timelike direction $N_\tau$ for two different lattice discretizations: 
HISQ/tree and Asqtad. The continuum limit is obtained for $N_\tau^{-2}\to 0$. Both figures taken from \cite{Bazavov:2011nk}.}
\label{fig:Tc}
\end{figure}

At $\mu_B/T = 0$, consistent results for the pseudo-critical temperature\index{pseudo-critical temperature} $T_c$ \cite{Bazavov:2011nk} and the equation of state of (2+1)-QCD \cite{HotQCD:2014kol,Borsanyi:2013bia} 
have been obtained from continuum extrapolated results of different lattice discretizations. The pseudo-critical temperature can be identified from the peak of chiral susceptibility\index{chiral!susceptibility} or other sensitive chiral observables.
This is illustrated in Fig.~\ref{fig:Tc}, where in the left panel results for the chiral susceptibility 
\begin{equation}
\chi_l = \frac{T}{V} \frac{\partial^2 \log Z}{\partial m_l^2}
\label{chiral_susceptibility}
\end{equation}
is shown as a function of the temperature. 
In the right panel of that figure the dependence of the pseudocritical temperature $T_c$ on the number of lattice sites $N_\tau$ in the timelike 
direction is shown for two different lattice discretizations: HISQ/tree and Asqtad. The continuum limit is obtained for $N_\tau^{-2}\to 0$, which corresponds to the limit of lattice spacing $a\to 0$, as the temperature is given by $T=1/(a N_\tau)$.

The most recent determination of the chiral cross-over\index{chiral!crossover} temperature \cite{HotQCD:2018pds} is based on the study of different chiral observables which in the continuum limit provide a consistent result of 
\begin{equation}
\label{eq:Tc}
  T_c = ( 156 \pm 1.5 ) \, {\rm MeV}. 
\end{equation}

The equation of state (EoS) of strong interaction matter, i.e. basic bulk thermodynamic observables like pressure ($P$), energy density ($\epsilon$) and entropy density\index{entropy} ($s$), provides important information for the understanding of QCD and its properties at high temperatures and densities. In recent years lattice QCD calculations have reached an accuracy that allows to obtain continuum extrapolated results at physical quark masses for these quantities.

\begin{figure}[htb]
 \includegraphics[width=0.48\textwidth]{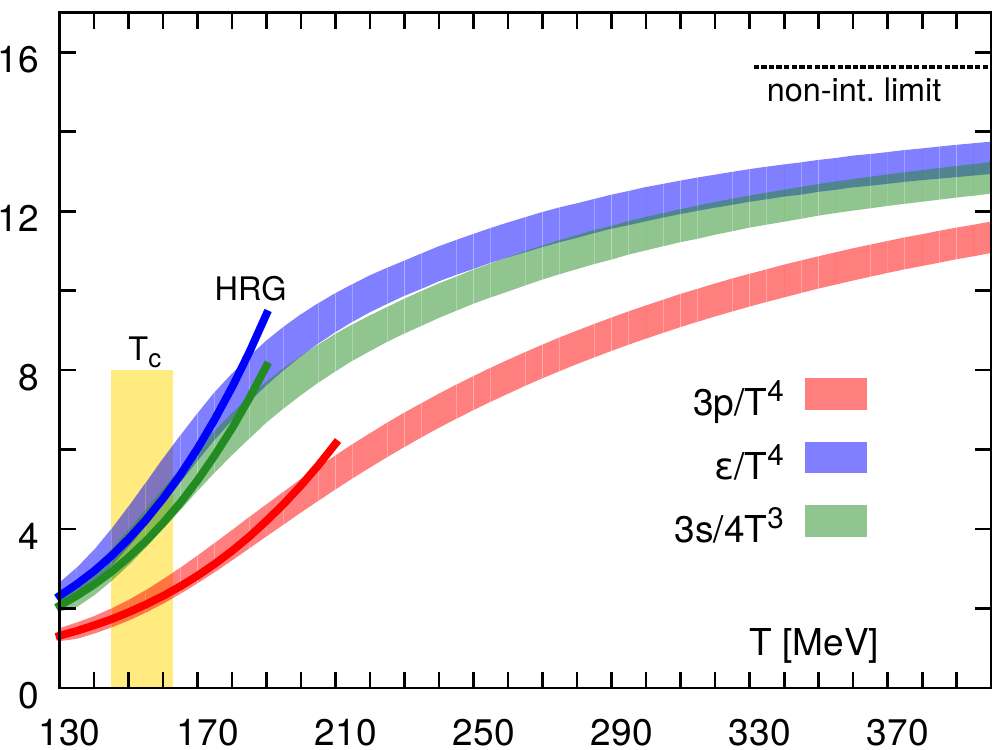}
 \hfill
 \includegraphics[width=0.48\textwidth]{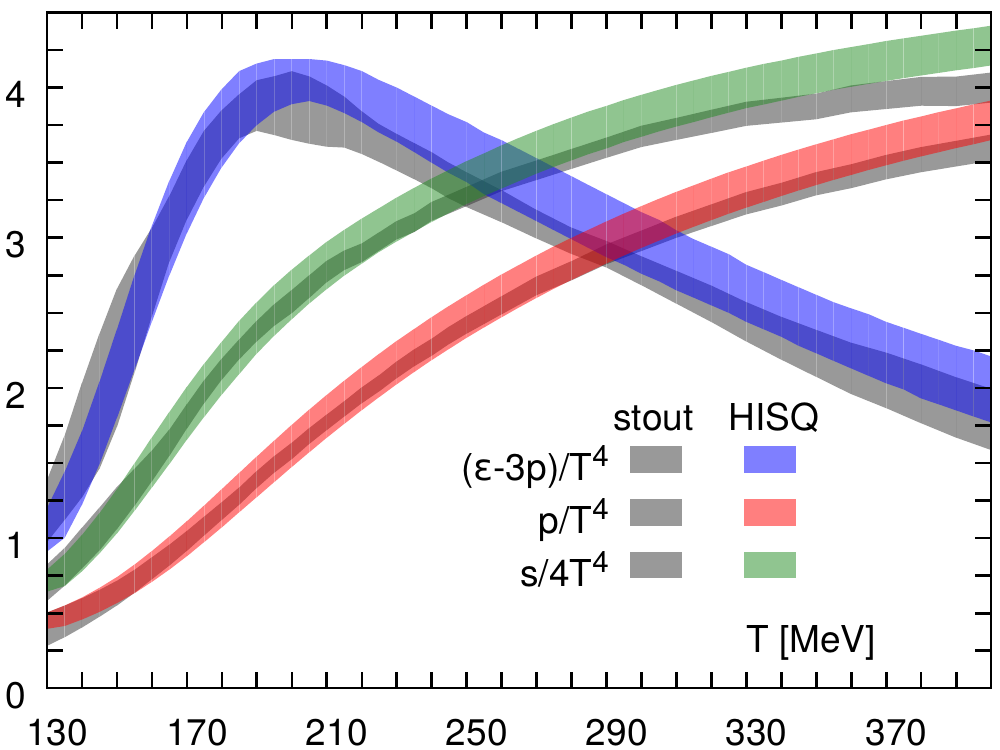}
\caption{Equations of state for (2+1)-flavor QCD with physical quark masses as functions of the temperature. 
Left panel: $3p/T^4$ (red band), $\varepsilon/T^4$ (blue band) and $3s/(4T^3)$ (green band). 
The solid lines of corresponding colors show the results for the hadron resonance gas (HRG) and the yellow rectangle 
indicates the pseudocritical temperature\index{pseudo-critical temperature} with its statistical error\index{statistical!error} band from Eq.~(\ref{eq:Tc}).
Right panel: Comparison of results for the temperature-dependence of the interaction measure $(\varepsilon-3p)$ (blue and grey bands), 
pressure $p$ (red and grey bands) and entropy $s/4$ (green and grey bands), all in units of $T^4$, for HISQ (colored) \cite{HotQCD:2014kol}
and stout (grey) \cite{Borsanyi:2013bia}
discretizations. Figures taken from \cite{HotQCD:2014kol}.}
\label{fig:EoS}
\end{figure}

The equation of state (EoS) can be obtained from the trace of the energy momentum tensor, $\Theta^{\mu \mu}$ (also called trace anomaly\index{trace anomaly} or interaction measure\index{interaction measure}), 
which is related to the pressure,
\begin{align}
  \frac{\Theta^{\mu \mu}(T)}{T^4} = \frac{\epsilon - 3 P}{T^4} = T \, \frac{d}{d T}\left(\frac{P}{T^4}\right)~.
\end{align}
On the lattice at finite temporal extent $N_t$, spatial extent $N_s$ and lattice spacing $a$ the temperature is given by $T=1/(a N_t)$, the volume by $V=(a N_s)^3$ and the equation can be expressed in terms of the derivative of the logarithm of the partition function $Z$ with respect to the lattice spacing,
\begin{align}
  \Theta^{\mu \mu} = \epsilon - 3 P = -\frac{T}{V} \, \frac{d \ln Z}{d \ln a}~.
\label{eq:trace_anomaly}
\end{align}
Although the partition function itself can not be calculated, the derivative of the logarithm of $Z$ with respect to the lattice spacing or equivalently to temperature in (\ref{eq:trace_anomaly}) leads to expressions in terms of expectation values of combinations the gauge action and light and strange quark chiral condensates\index{chiral!condensate}, see for instance \cite{Cheng:2007jq,HotQCD:2014kol,Karsch:2001cy}. The remaining thermodynamic functions can be derived from $\Theta^{\mu \mu}$. The pressure $P$ can be determined using the so called integral method up to an integration constant at a temperature $T_0$ by integrating the trace anomaly\index{trace anomaly},
\begin{align}
\frac{P(T)}{T^4} - \frac{P(T_0)}{T^4} = \int_{T_0}^T \mathrm{d}T' 
\frac{\epsilon - 3 P}{T'^5}~,
\end{align}
which allows to reconstruct the energy density through (\ref{eq:trace_anomaly}) and the entropy\index{entropy} density as $s/T^3=(\epsilon+P)/T^4$.

Continuum extrapolated results for pressure, energy density and entropy density are shown in Fig.~\ref{fig:EoS} for two different lattice calculations using different lattice discretizations. 
Although on finite lattices these discretization effects can be different, 
from the right panel of that figure we can conclude that the results from the hotQCD collaboration using the HISQ fermion discretization \cite{HotQCD:2014kol} and from the Budapest-Wuppertal collaboration using the stout fermions 
\cite{Borsanyi:2013bia} show a good agreement in the continuum.

In the high temperature limit bulk thermodynamic observables are expected to
approach their free gas values, i.e. the non-interacting Stefan-Boltzmann limit. The deviations from this limit seen in Fig.~\ref{fig:EoS} indicate that QCD matter is still strongly interacting rather than a weekly interacting gas at the highest temperatures shown here.
Nevertheless, the strong increase of the thermodynamic quantities around the crossover region, indicated by the yellow band, shows the liberation of the quark and gluon degrees of freedom in the high temperature phase, while at temperatures up to the crossover region a hadron resonance gas model, which will be described in more detail in the next section, can be used to describe the thermodynamics rather well. 

The comparison with the hadron resonance gas (HRG) in the left panel shows that
the HRG model, using all known hadronic resonances from the PDG compilation \cite{ParticleDataGroup:2020ssz}, describes
the QCD equation of state quite well up to the crossover region.
The QCD results lie systematically above the HRG ones for $T<160$ MeV, which is a first indication that additional resonances which are not listed in the PDG, i.e. which are not yet experimentally observed may contribute. This will be discussed in the next section in more detail. At higher temperatures the QCD thermodynamics can no longer be described by a hadron resonance gas as the hadronic degrees of freedom cease to be present and a strongly interacting gas of fermionic and gluonic degrees of freedom take over.
QCD is quite different from HRG thermodynamics at $T>160$ MeV.

\subsection{HRG and missing strange baryons}\label{sec:hrg_strange}

In the previous section we have seen that a hadron resonance gas model provides a good approximation of the QCD thermodynamics up to the crossover region. 
In the HRG model, hadrons are treated as an uncorrelated free gas and the thermodynamic pressure is simply given by the sum of their partial pressure,
\begin{eqnarray}
    P_{\rm HRG} &\approx& \sum_{\rm all \, hadrons} \, \frac{T^4}{2 \pi^2}  g_h  (\frac{m_h}{T})^2   K_2(m_h/T) 
    \cosh [ B_h \hat{\mu}_B + Q_h \hat{\mu}_Q + S_h \hat{\mu}_S ]
\end{eqnarray}
Here $g_h$ is the degeneracy factor of hadrons of mass $m_h$ and $\hat{\mu}_X=\mu_X/T$ the chemical potential with respect to the quantum numbers for baryon number, electric charge and strangeness, $X=B, Q, S$.
The list of hadrons included can be the states listed in the particle data group (PDG-HRG), i.e. which have been experimentally observed, or in addition those calculated within a quark model (QM-HRG), 
Where the latter may include additional strange mesons and baryons predicted by the quark model \cite{Capstick:1986bm,Ebert:2009ub}. A large number of additional resonances has also been
identified in lattice QCD calculations \cite{Edwards:2012fx}.

\begin{figure}[t]
\begin{center}
 \includegraphics[height=0.6\textwidth]{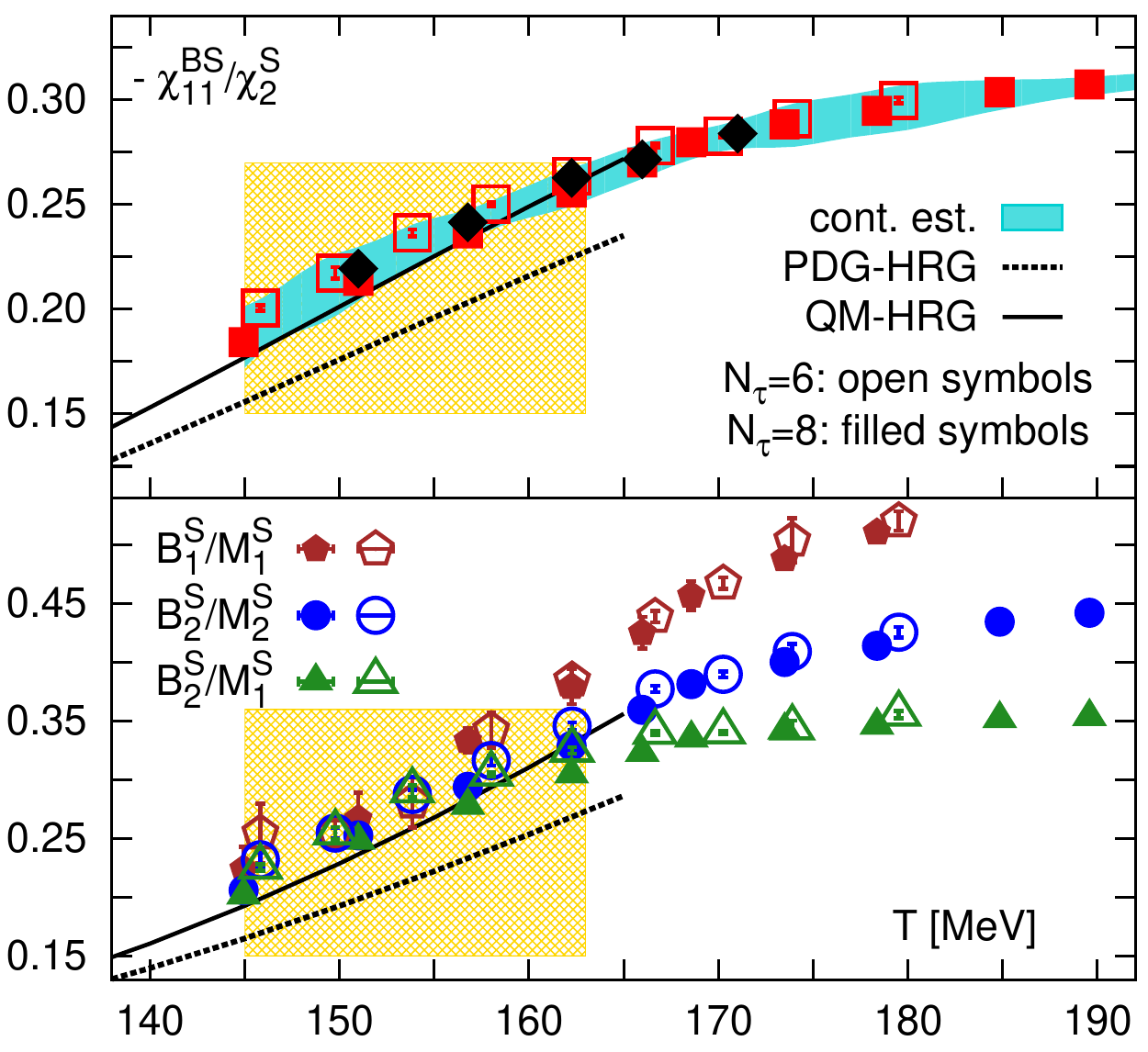}
\caption{Generalized susceptibilities and their ratios relevant for studying strangeness in the interacting system \cite{Bazavov:2014xya}.
The lattice QCD results are compared to two versions of the hadron resonance gas model (HRG). See the text for more details.
Taken from \cite{Bazavov:2014xya}.
}
\label{fig3}
\end{center}
\end{figure}


By comparing the HRG models to the lattice calculation, one can obtain important information about the hadronic content of the thermal medium\index{thermal!medium}.
For example, lattice QCD suggests the existence of additional strange baryons beyond the known ones listed  in the PDG table. 
This can be inferred from the study of the relevant generalized susceptibilities, which are defined as Taylor coefficients of the pressure expanded in the chemical potentials of  conserved charges, i.e. baryon number $B$, electric charge $Q$ and strangeness $S$,
\begin{align}
\frac{P(T)}{T^4} = \sum_{i,j,k=0}^{\infty}
\frac{1}{i!j!k!}
\chi_{ijk}^{BQS}(T)
\left(\frac{\mu_B}{T}\right)^i
\left(\frac{\mu_Q}{T}\right)^j
\left(\frac{\mu_S}{T}\right)^k
\label{eq:taylor}
\end{align} 
The generalized susceptibilities can be calculated at vanishing chemical potential by
\begin{equation}
  \chi_{ijk}^{BQS}(T) = \frac{\partial^{i+j+k} P(T,\hat{\mu}_B,\hat{\mu}_Q,\hat{\mu}_S)/T^4}{\partial \hat{\mu}_B^i \partial \hat{\mu}_Q^j \partial\hat{\mu}_S^k} \bigg|_{\vec{\mu}=0} .
\label{eq:suscept}
\end{equation}
These susceptibilities can be trivially computed in the HRG models by taking relevant derivatives. 
Any trivial volume dependence in the susceptibilities can be absorbed by taking ratios of two different susceptibilities of the same order.
In particular, we will first look at the following ratio
\begin{align}
  \frac{\chi_{11}^{\rm BS}}{\chi_{2}^{\rm S}},
  \label{eq:ratio}
\end{align}
which provides the correlation of net strangeness with net baryon number fluctuations normalized to the second cumulant of net strangeness fluctuations.
This ratio is one of the sensitive probes of strangeness. Within the HRG model, the numerator is given by the strange baryon density while the denominator is dominated by strange mesons. 
Results of this ratio from lattice calculations with almost physical quark masses ($m_s/m_l=20$) using the HISQ action \cite{Bazavov:2014xya} can be found in Fig.~\ref{fig3}.
Also shown are two linearly independent pressure-observables for
the open strange meson ($M_1^S, M_2^S$) and baryon ($B_1^S, B_2^S$) partial pressures, which are defined in \cite{Bazavov:2014xya}.
The agreement of all these strangeness sensitive pressure observables up to the crossover region shows that a description of QCD thermodynamics in terms of an uncorrelated gas of hadrons is valid
till the chiral crossover\index{chiral!crossover} region (yellow band), while the deviations at higher temperatures are an indication for the liberation of strange degrees of freedom. 
In all observables shown here it is found that the QM-HRG scheme, which contains some so-far undiscovered strange baryonic states, gives a better description of the lattice result for these observables and that these states give a sensible contribution to the thermodynamics in the strangeness sector. 

\begin{figure}[t]
\begin{center}
 \includegraphics[height=0.45\textwidth]{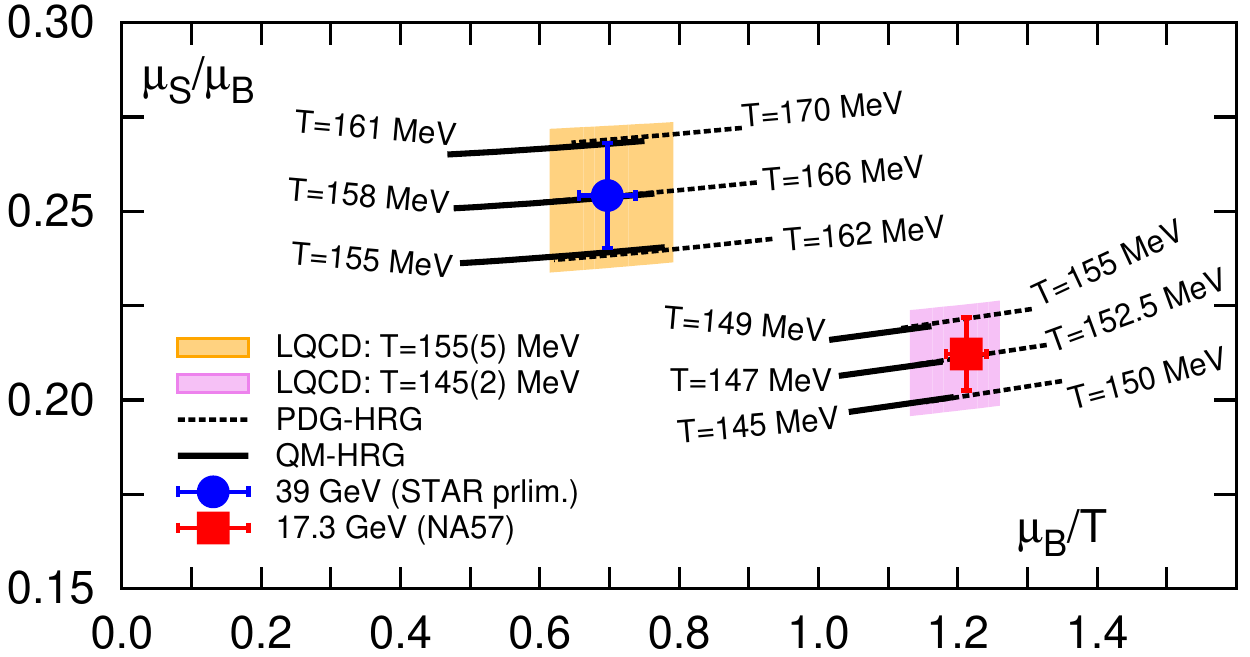}
  \caption{Freeze-out parameters imposing strangeness neutrality. Experimental results for NA57 and STAR are compared to lattice QCD results and predictions of PDG-HRG (dotted lines) and QM-HRG (solid lines). Taken from \cite{Bazavov:2014xya}.}
\label{fig4}
\end{center}
\end{figure}

\subsection{Strangeness freeze-out\index{freeze-out!strangeness}}
Freeze-out is the heuristic concept that the evolution of the system, e.g. by expansion, proceeds via quasi-equilibrium states defined by a temperature and a set of chemical potentials until at one time step the rate for the equilibration processes drops below the expansion rate and the particle distributions do not change upon further expansion and cooling of the system - they are 'frozen out'. The freeze-out of the momentum distribution is called "kinetic freeze-out" or "thermal freeze-out"\index{freeze-out!kinetic/thermal}, while the freezing of the distribution of particle species 
is called "chemical freeze-out"\index{freeze-out!chemical}. 
Since the freeze-out\index{freeze-out} is a sudden process, the particle distributions at the very moment of freeze-out are 
still described by the parameters of equilibrium thermodynamics. 
A comparison between experimental results and theoretical predictions based on either model descriptions
(e.g., hadron resonance gas model) or ab-initio lattice QCD data shows that the concept of a sudden freeze-out 
is very successful \cite{Andronic:2017pug}.
For further details, see the chapter on statistical hadronization\index{statistical!hadronization} by Radoslaw Ryblewski.

We will now discuss how lattice QCD results of suitable ratios of generalized susceptibilities compared to experimental results can be used to determine the freeze-out conditions for strangeness.

The system of initial nuclei in a heavy ion collision is net strangeness neutral. This constraint,
\begin{align}
\langle n_S \rangle = 0
\end{align}
implies that the strangeness chemical potential, $\mu_S$, can be expressed in terms of the temperature, $T$ and the baryon chemical potential, $\mu_B$. To leading order in $\mu_B$, it can be shown that
\begin{align}
  \left( \frac{\mu_S}{\mu_B} \right) \approx - \frac{\chi_{11}^{\rm BS}}{\chi_{2}^{\rm S}} - \frac{\chi_{11}^{\rm QS}}{\chi_{2}^{\rm S}} \frac{\mu_Q}{\mu_B},
\end{align}
where the second term provides a small correction involving the electric charge chemical potential, $\mu_Q$, and higher order terms in $\mu_B/T$ are small for values of $\mu_B$ smaller than 200~MeV.

The relative yields of strange anti-baryons $(\bar{H}_S)$ to strange baryons $(H_S)$ can be used to determine the freeze-out parameters\index{freeze-out!parameters} $\mu^f_B/T^f$ and $\mu^f_S/\mu^f_B$ via
\begin{align}
  R_H \equiv \frac{\bar{H}_S}{H_S} = e^{-2(\mu^f_B/T^f)(1-(\mu^f_S/\mu^f_B)\vert S \vert)}.
\end{align}
Experimental results for these strangeness freeze-out parameters for NA57 \cite{NA57:2004nxc} and STAR \cite{Zhao:2013yza} are shown in Fig.~\ref{fig4}. These results can be compared to lattice QCD results and predictions for the different hadron resonance gas models. Varying the temperature in these calculations and matching the values to experiment allows to determine the freeze-out temperature. While the QM-HRG predictions and the lattice QCD results are in good agreement, the freeze-out temperatures of PDG-HRG are larger.  
This again shows the sensitivity of the HRG to the strangeness particle content and their relevance to the thermodynamics in the vicinity of the QCD crossover. 

\subsection{Taylor expansion in chemical potential}

In this section, we discuss lattice QCD results on the equation of state (EoS) at finite baryon density based on a Taylor expansion in the chemical potentials of conserved charges, i.e. in the baryon, strangeness and electric charge chemical potentials. 
The Taylor expansion of the pressure Eq.~(\ref{eq:taylor}) and the corresponding expansion coefficients defined by the generalized susceptibilities, Eq.~(\ref{eq:suscept}) have been introduced already in section~\ref{sec:hrg_strange}.
For $\mu_S = \mu_Q = 0$, the expansion coefficients are determined solely by the fluctuations of baryon charges at $\mu_B = 0$ and the expansion of the pressure, $P(T,\mu_B)$, and the net baryon-number density, $n_B$, becomes
\begin{align}
\begin{split}
\frac{P(T,\mu_B)-P(T,0)}{T^4} &= \sum_{n=1}^\infty \frac{\chi_{2n}^{B}(T)}{(2n)!} \left(\frac{\mu_B}{T}\right)^{2n}   \\
  &= \frac{1}{2} \chi_2^B(T) \hat{\mu}_B^2 \left( 1+ \frac{1}{12} \frac{\chi_{4}^B(T)}{\chi_2^B(T)} \hat{\mu}_B^2 + \frac{1}{360}\frac{\chi_{6}^B(T)}{\chi_2^B(T)} \hat{\mu}_B^4+\; ... \right),
  \end{split}
\end{align} 
\begin{align}
\begin{split}
  \frac{n_B}{T^3} = \frac{\partial P/T^4}{\partial \hat{\mu}_B} &= \sum_{n=1}^\infty \frac{\chi_{2n}^{B}(T)}{(2n-1)!}\hat{\mu}_B^{2n-1} \\
  &= \chi_2^B(T) \hat{\mu}_B \left( 1+ \frac{1}{6} \frac{\chi_{4}^B(T)}{\chi_2^B(T)} \hat{\mu}_B^2 + \frac{1}{120}\frac{\chi_{6}^B(T)}{\chi_2^B(T)} \hat{\mu}_B^4+\; ... \right).
\end{split}
\end{align} 
%
\begin{figure}[t]
 \includegraphics[height=0.34\textwidth]{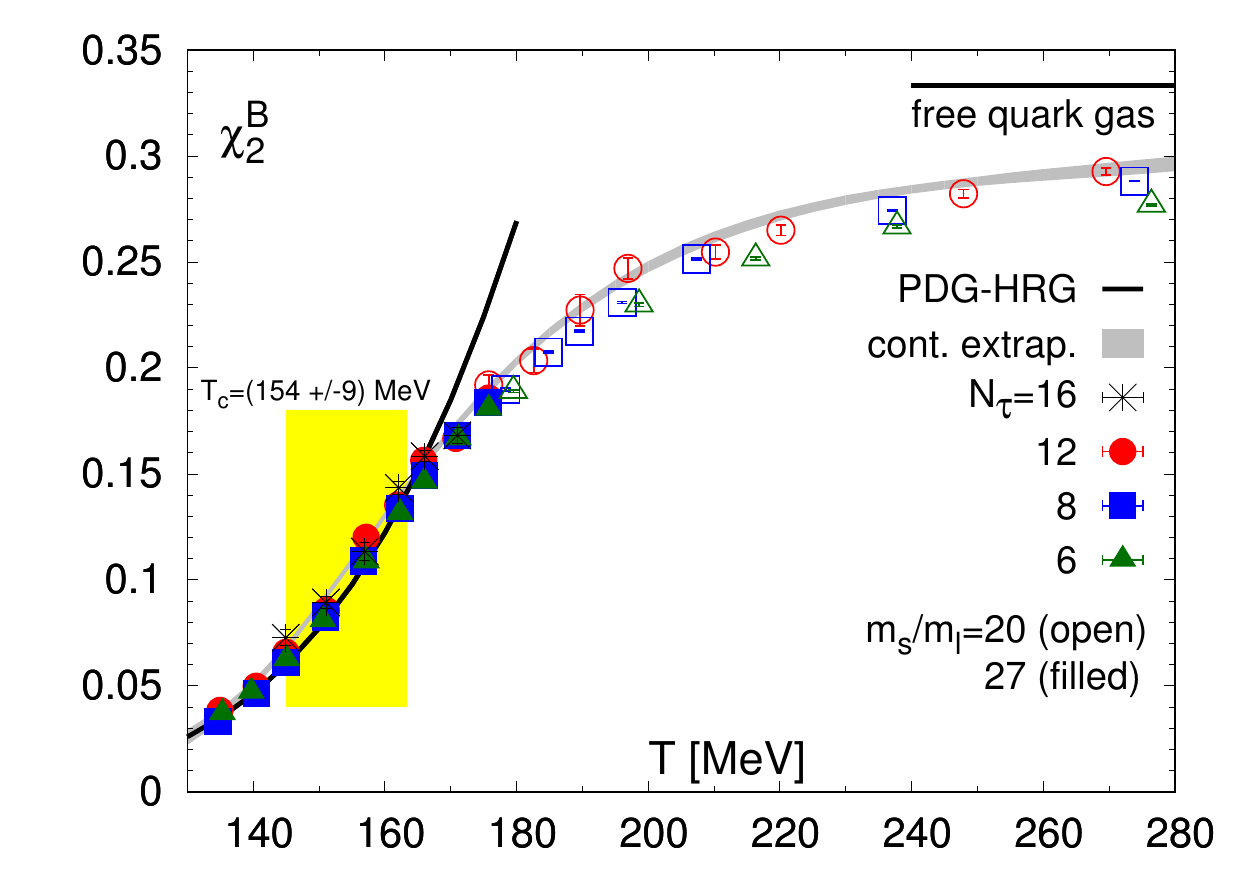}
\hfill
 \includegraphics[height=0.34\textwidth]{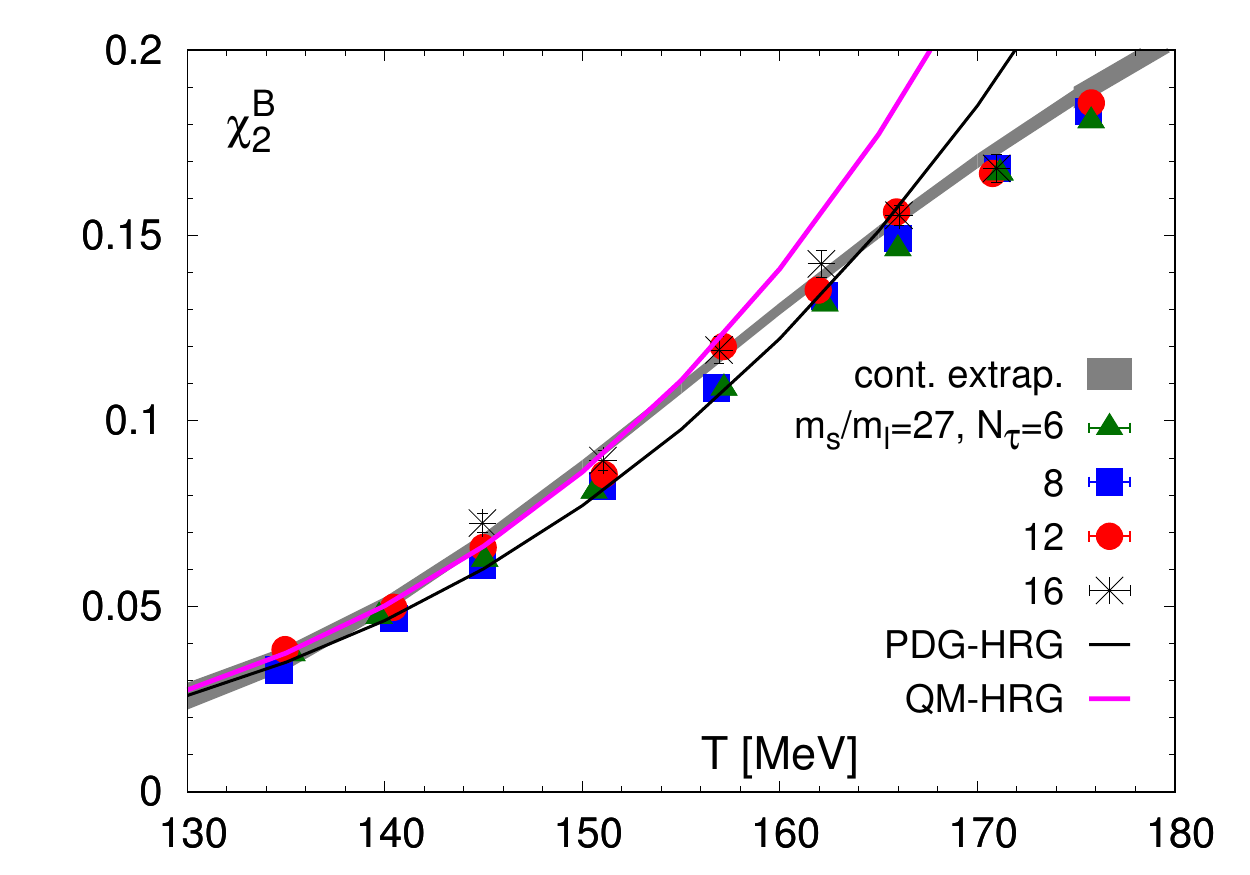}\\
  \includegraphics[width=0.48\textwidth]{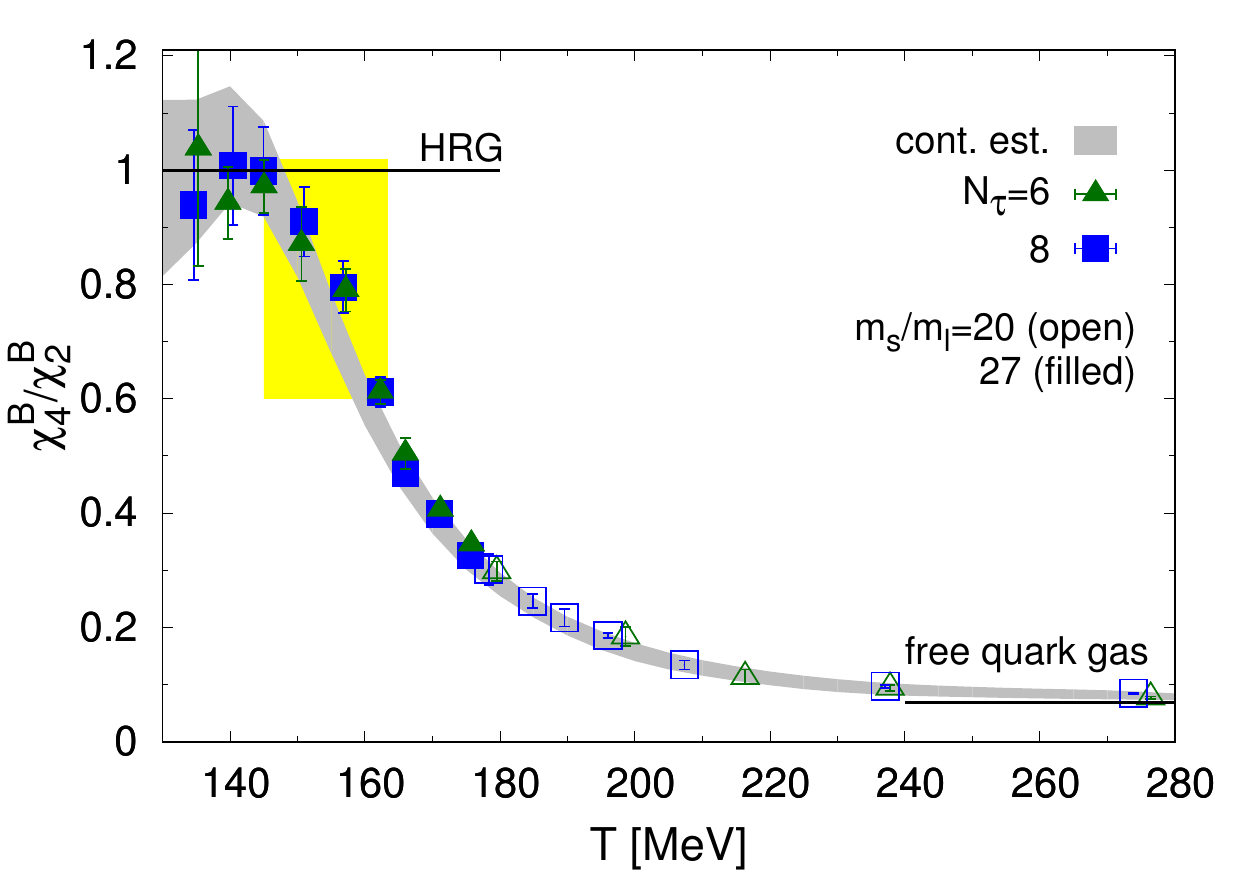}
\hfill
 \includegraphics[width=0.47\textwidth]{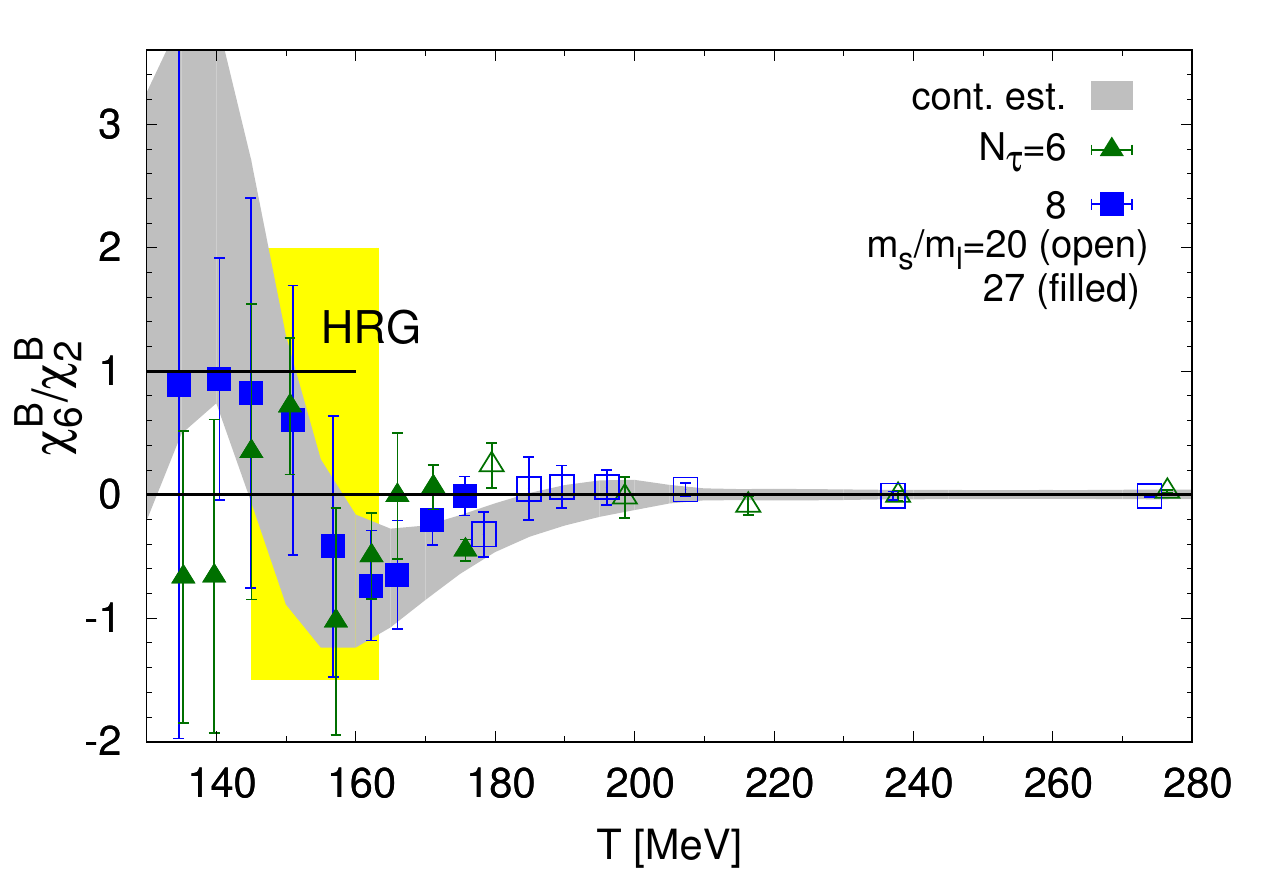}

  \caption{Taylor expansion coefficient $\chi_2^B$ and ratios $\chi_4^B/\chi_2^B$ and $\chi_6^B/\chi_2^B$ as a function of the temperature under the condition $\mu_Q = \mu_S = 0$~\cite{Bazavov:2017dus}.}
\label{fig5a}
\end{figure}
%
Lattice QCD results and continuum estimates of the first Taylor expansion coefficients from \cite{Bazavov:2017dus} are shown in Fig.~\ref{fig5a}. These correspond to the variance of the net baryon number distribution $\chi_2^B$. 
In Fig.~\ref{fig5b}~(left) the ratio of fourth and second order cumulants of net-baryon number 
fluctuations, that corresponds to the kurtosis times variance $\kappa_B \sigma_B^2 = \chi_4^B/\chi_2^B$, is shown as a function of the temperature.
Fig.~\ref{fig5b}~(right) shows the ratio of fourth and second order cumulants.
%
\begin{figure}[t]
 \hspace*{-0.38cm}
 \includegraphics[width=0.58\textwidth]{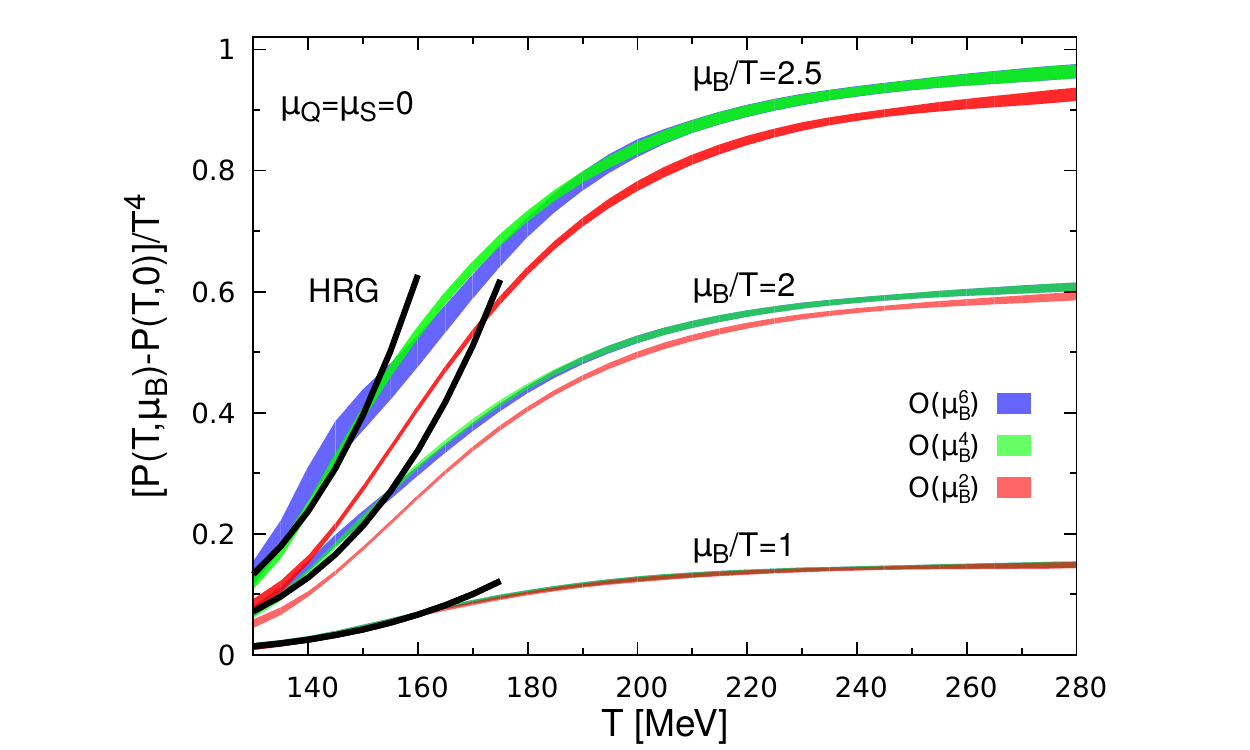}
 \hspace*{-1.3cm}
 \includegraphics[width=0.58\textwidth]{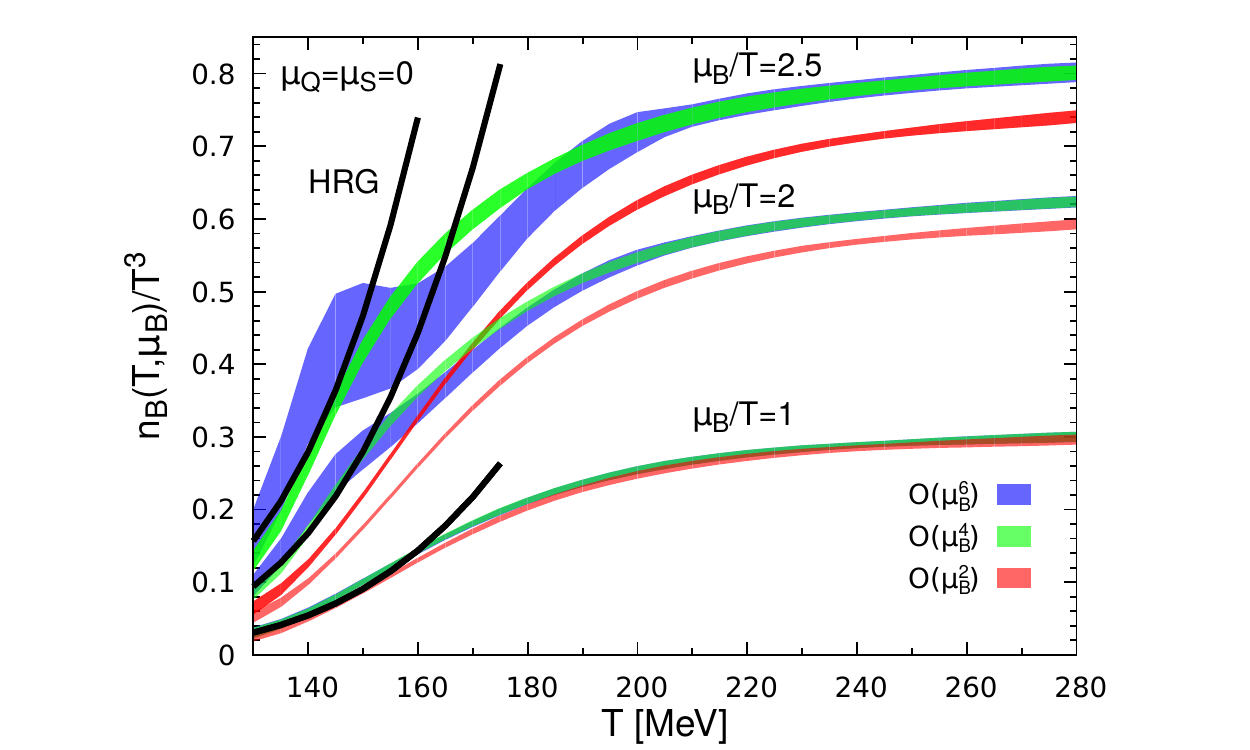}
  \caption{The $\mu_B$-dependent contribution to the pressure (left) and the baryon-number density (right) for vanishing
electric charge and strangeness chemical potentials, from \cite{Bazavov:2017dus}.}
\label{fig5b}
\end{figure}

\subsection{Equation of state in a strangeness neutral system}

\begin{figure}[thb]
 \includegraphics[width=0.48\textwidth]{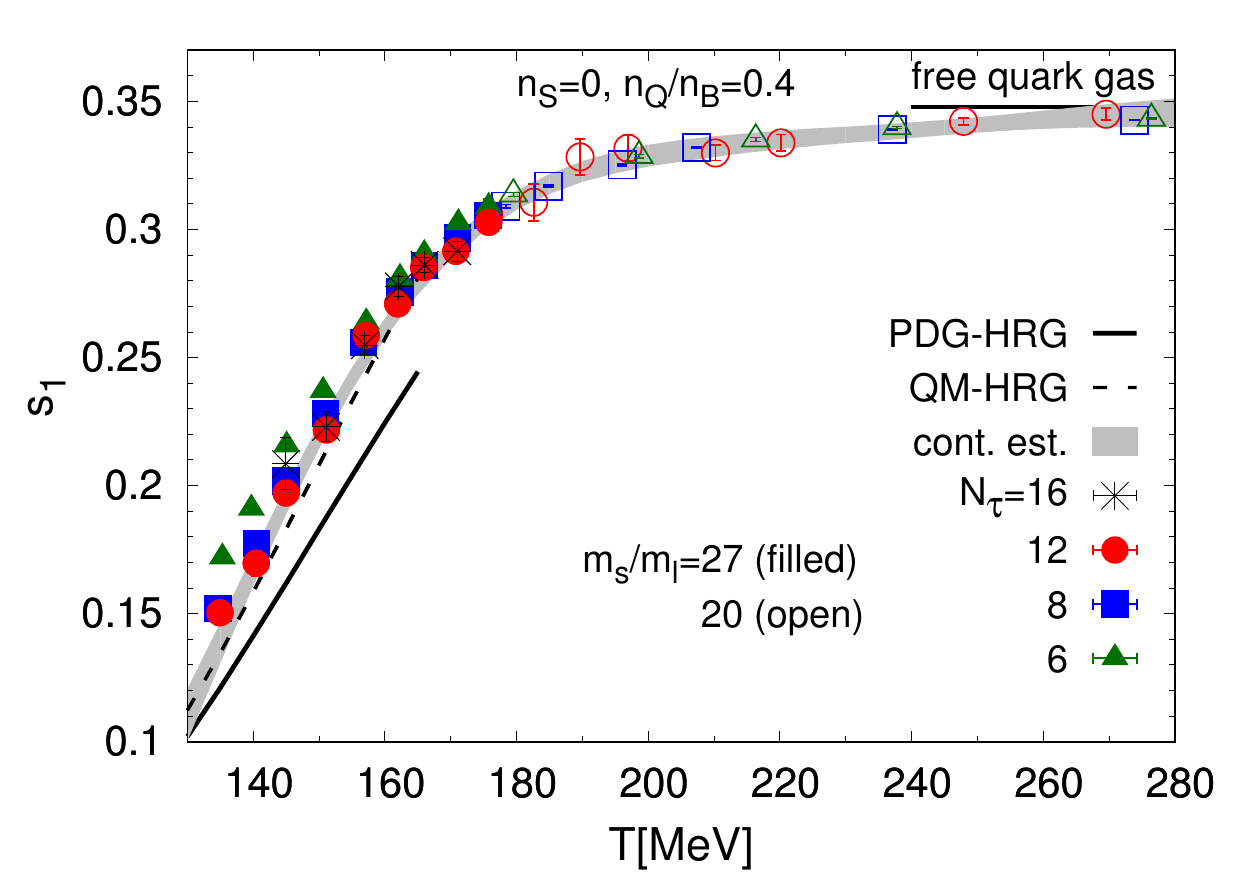}
 \hfill
 \includegraphics[width=0.48\textwidth]{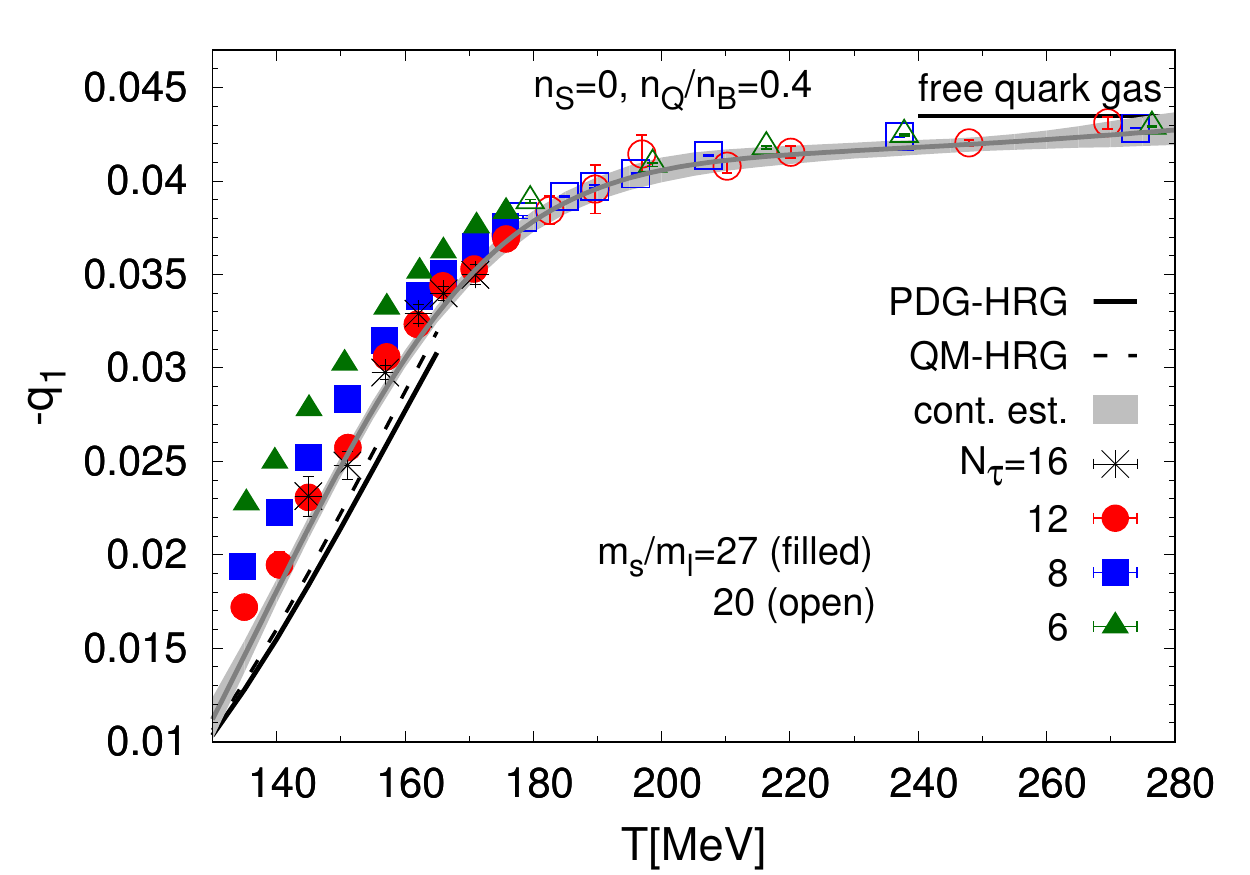}
  \caption{Leading order expansion coefficients, $s_1(T)$ and $-q_1(T)$, for the expansion of $\hat{\mu}_S(T,\mu_B)$ and $\hat{\mu}_Q(T,\mu_B)$ with respect to $\hat{\mu}_B$. Taken from \cite{Bazavov:2017dus}.}
\label{fig6a}
\end{figure}

\begin{figure}[thb]
\centering{
 \includegraphics[width=0.48\textwidth]{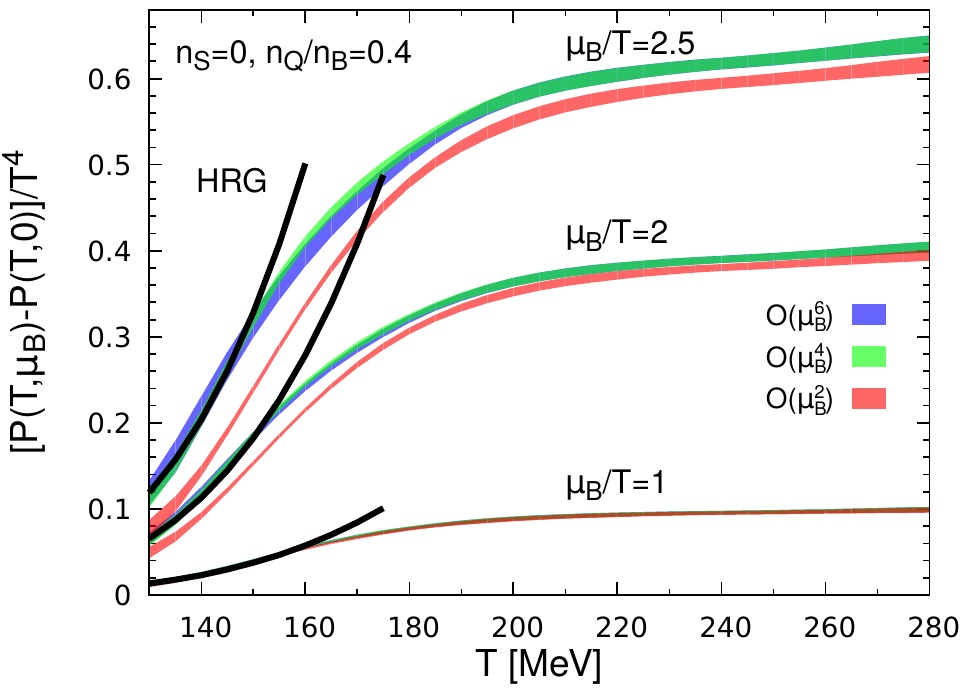}
 \hfill
 \includegraphics[width=0.42\textwidth]{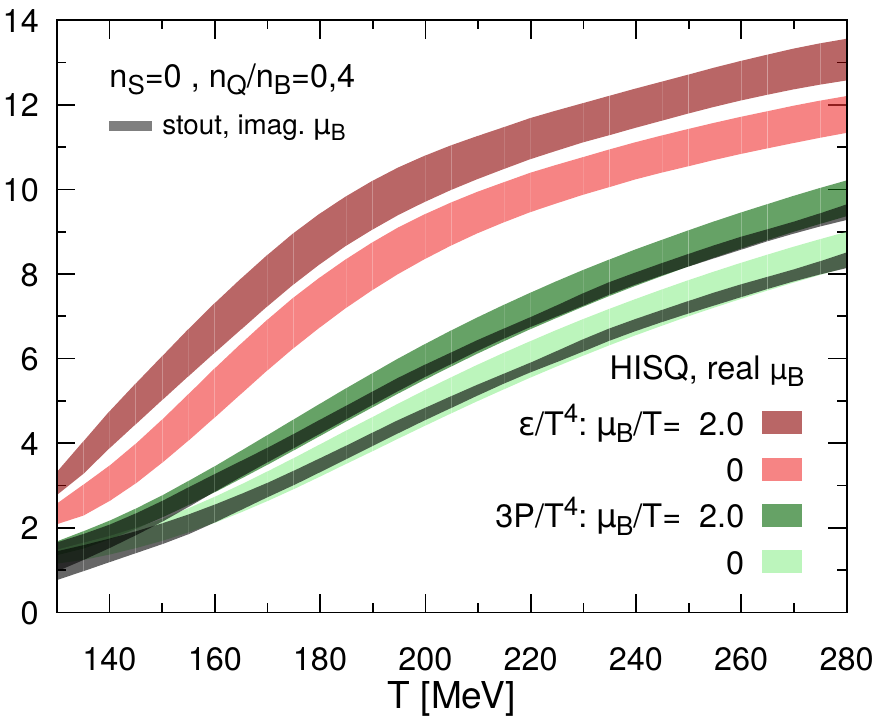}
 }
  \caption{Left: The $\mu_B$-dependent contribution to the pressure for a strangeness neutral system with $r=0.4$ for some values of the baryon chemical potential, $\mu_B/T$. Right: The total energy density and three times the pressure of \cite{Bazavov:2017dus} compared to stout results of \cite{Guenther:2017hnx}. Both plots are taken from \cite{Bazavov:2017dus}.}
\label{fig6b}
\end{figure}

To match to the conditions of the heavy ion collision, the constraint of strangeness neutrality and fixed ratio of electric charge to baryon-number density should be imposed,
\begin{align}
  \begin{split}
    \langle n_S \rangle &= 0,~~    n_Q/n_B = r.
  \end{split}
\end{align}
As previously discussed, this allows the determination of $\hat{\mu}_Q$ and $\hat{\mu}_S$ in terms of $(T, \mu_B)$, here as expansions in $\hat{\mu}_B$
\begin{align}
  \hat{\mu}_Q(T,\mu_B) &= q_1(T) \hat{\mu}_B + q_3(T) \hat{\mu}_B^3 + q_5(T) \hat{\mu}_B^5 + \cdots \\
  \hat{\mu}_S(T,\mu_B) &= s_1(T) \hat{\mu}_B + s_3(T) \hat{\mu}_B^3 + s_5(T) \hat{\mu}_B^5 + \cdots.
\end{align}
These expansion parameters can be determined from lattice QCD calculations at $\mu_B/T=0$. The results for the leading order expansion coefficients, 
$s_1(T)$ and $-q_1(T)$, for the expansions of $\hat{\mu}_S(T,\mu_B)$ and $\hat{\mu}_Q(T,\mu_B)$ with respect to $\hat{\mu}_B$ can be found in Fig.~\ref{fig6a}.
The result of the total energy density and three times the pressure of \cite{Bazavov:2017dus} compared to stout results of \cite{Guenther:2017hnx} are shown in Fig.~\ref{fig6b}. 
The calculations with HISQ and stout actions provide consistent results in the continuum for $\mu_B/T \leq 2$, which correspond to a collision energy 
$\sqrt{s_{NN}} \geq 14.5 $ GeV. 
%
\begin{figure}[!thb]
\centering{
 \includegraphics[width=0.65\textwidth]{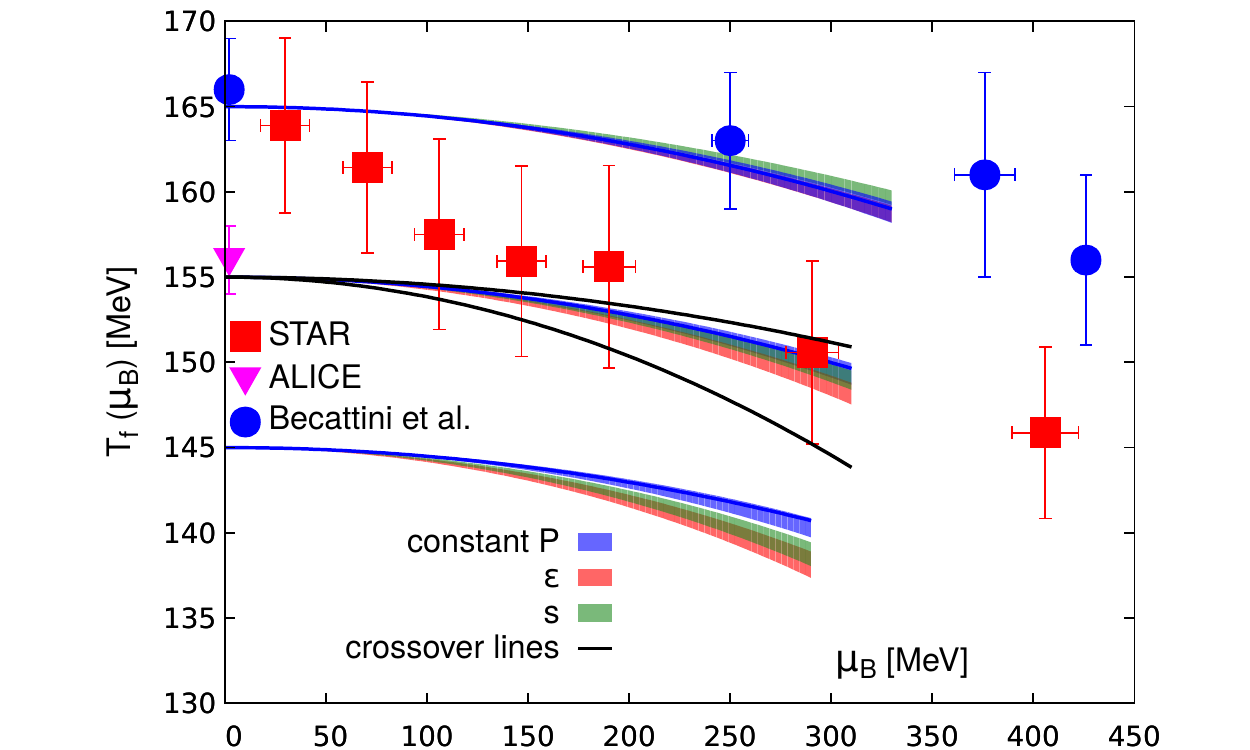}
  \caption{Lines of constant physics determined by lattice QCD. See the text for a more detailed explanation. Taken from \cite{Bazavov:2017dus}. }
  }
  \label{fig7a}
\end{figure}

The thermal conditions at the chemical freeze-out\index{freeze-out!chemical} (hadronization) in heavy ion collisions are characterized by lines of constant
pressure, energy and entropy\index{entropy} density in the $T-\mu_B$ plane.
For different observable $f(T,\mu_B)$, like pressure $P$, energy density $\epsilon$ or entropy density $s$, these lines can be parameterized for small $\mu_B$ as an expansion in $\mu_B/T_0$ where $T_0$ is the corresponding temperature at vanishing chemical potential,    
\begin{equation}
T_f(\mu_B) = T_0 \left(1 - \kappa_2^f \left( \frac{\mu_B}{T_0}\right)^2- \kappa_4^f \left( \frac{\mu_B}{T_0}\right)^4 \right) ~.~
\end{equation}
Around $T_c$ consistent results for the three observables  were estimated in \cite{Bazavov:2017dus} 
with $0.0064 \leq \kappa_2^{P} \leq 0.0101$, $0.0087 \leq \kappa_2^{\varepsilon} \leq 0.012$ and $0.0074 \leq \kappa_2^{s} \leq 0.011$.
These results correspond well to that of the chiral crossover\index{chiral!crossover} line for the QCD transition for which a range of
$0.0066 \leq \kappa_2^{c} \leq 0.020$ is consistent with results of various groups using different methods \cite{Kaczmarek:2011zz,Endrodi:2011gv,Bonati:2015bha,Bellwied:2015rza,Cea:2015cya}.
Fig.~\ref{fig7a} shows this crossover line and lines of constant physics for the three observables for three values of $T_0=145, 155$ and $165$ MeV. Also shown are freeze-out
temperatures determined by the STAR Collaboration in the beam energy scan program (BES) at RHIC\index{RHIC} \cite{Das:2014qca} and the ALICE Collaboration at the LHC\index{LHC} \cite{Floris:2014pta}. In addition hadronization temperatures obtained by comparing experimental data on particle yields with a
hadronization model calculation \cite{Becattini:2016xct} are also shown.
For a recent update on this analysis including various other conditions,
also in the strangeness ($\mu_S$), 
electric charge ($\mu_Q$) and isospin ($\mu_I$) directions
see \cite{HotQCD:2018pds}.

\subsection{Melting and abundance of open charm hadrons}
\begin{figure}[htbp]
\centering{
 \includegraphics[width=0.45\textwidth]{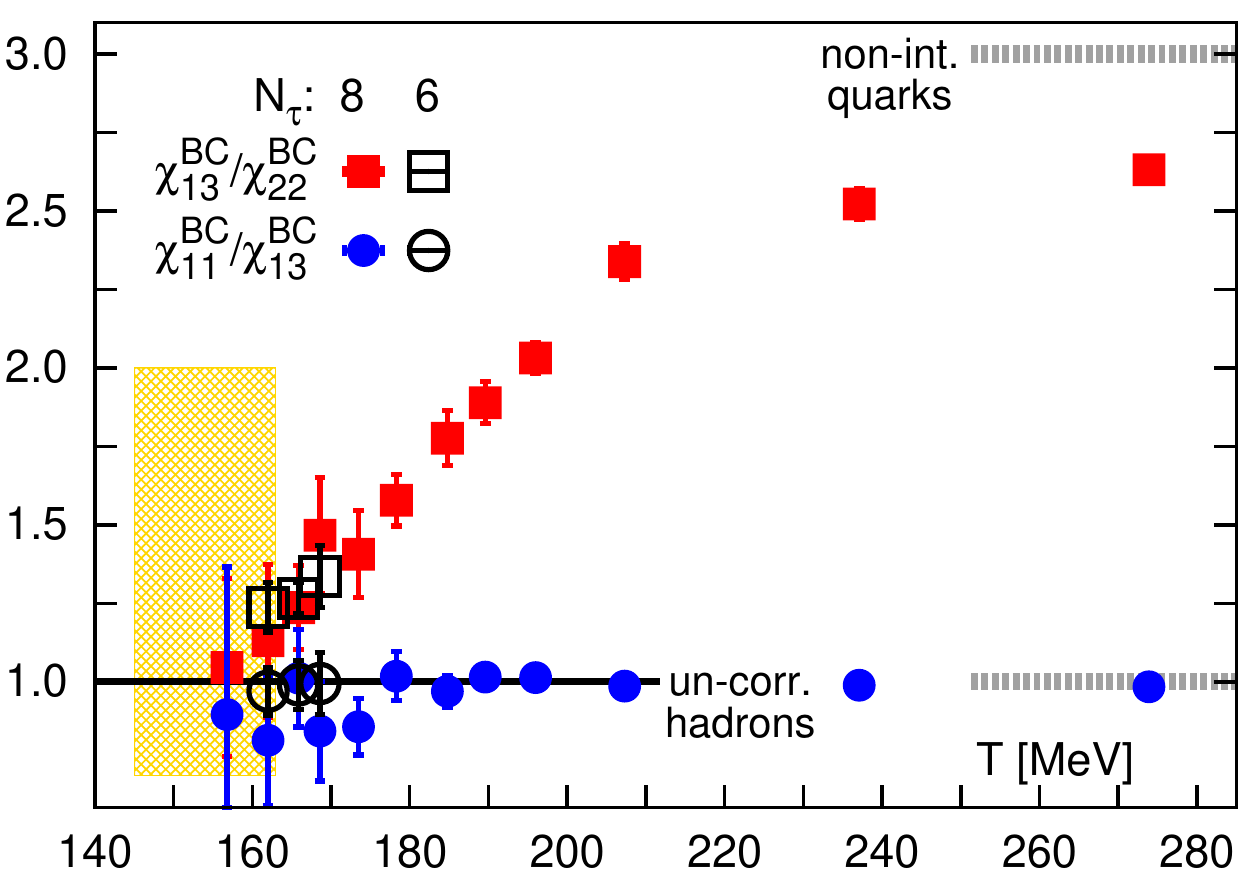}
 \includegraphics[width=0.45\textwidth]{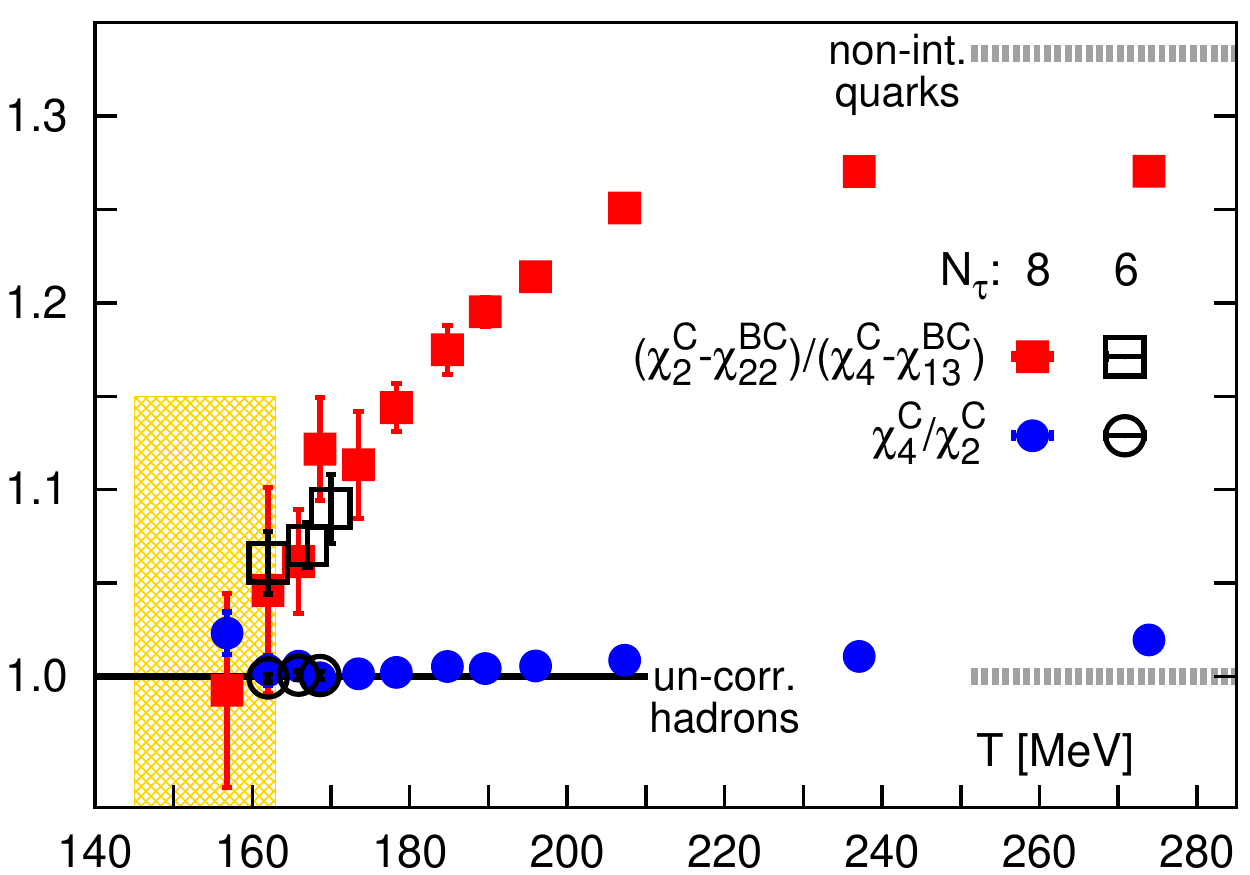}
 \caption{Left: Ratios of fourth order baryon-charm (BC) correlations that are sensitive to charmed baryons.
 Right: Combinations of charm fluctuations and baryon-charm correlations that only receive contributions from open charm mesons. Taken from \cite{Bazavov:2014yba}.}
  }
  \label{fig:charm1}
\end{figure}
The generalized susceptibilities defined in Eq.~(\ref{eq:suscept}) can also be extended to include charm degrees of freedom, i.e. including a charm chemical potential, $\hat\mu_C$, one can define 
\begin{equation}
  \chi_{ijkl}^{BQSC}(T) = \frac{\partial^{i+j+k+l} P(T,\hat{\mu}_B,\hat{\mu}_Q,\hat{\mu}_S,\hat{\mu}_C)/T^4}{\partial \hat{\mu}_B^i \partial \hat{\mu}_Q^j \partial\hat{\mu}_S^k \partial\hat{\mu}_C^l} \bigg|_{\vec{\mu}=0} .
\label{eq:susceptC}
\end{equation}
These can be used to analyze ratios of cumulants of correlations between net charm fluctuations and net-baryon number fluctuations ($BC$-correlations) as well as cumulants of net charm fluctuations ($\chi_n^C$) \cite{Bazavov:2014yba}. A comparison of these fluctuation observables to the HRG model can be used to determine the validity range of an uncorrelated hadron resonance gas model description of the open charm sector of QCD. Deviations from a hadron resonance gas model may then indicate the melting of open charm degrees of freedom. In the charmed baryon sector of thermodynamics, $\vert C \vert = 1$ baryons give the dominant contribution. In a good approximation, this leads to a simple relation of the $BC$-correlations in the hadronic phase,
\begin{equation}
\chi_{nm}^{BC} \simeq \chi_{11}^{BC} \;\; ,\;\ n+m > 2\; {\rm and~even}\ .
\label{eq:BC}
\end{equation}
While the ratio $\chi_{11}^{BC}/\chi_{13}^{BC}$ indeed is unity not only in the hadronic phase but also for an uncorrelated charmed quark gas, higher order ratios like $\chi_{13}^{BC}/\chi_{22}^{BC}$ are unity only in the hadronic phase but deviate from unity when 
an uncorrelated gas of charmed baryons is no longer a good description of the system.
These two ratios that are sensitive to the charmed baryon sector are shown in Fig.~\ref{fig:charm1}~(left).
Suitable combinations of charmed susceptibilities can be used to calculate observables that are sensitive to the open charm meson sector. Two of these are shown in Fig.~\ref{fig:charm1}~(right). 
The results in both sectors show 
that a description of a gas of uncorrelated charmed hadrons breaks down in or just above the chiral crossover region, therefore 
indicating that open charm degrees of freedom start to dissolve already close to the chiral crossover.

\begin{figure}[thbp]
\centering{
 \includegraphics[width=0.35\textwidth]{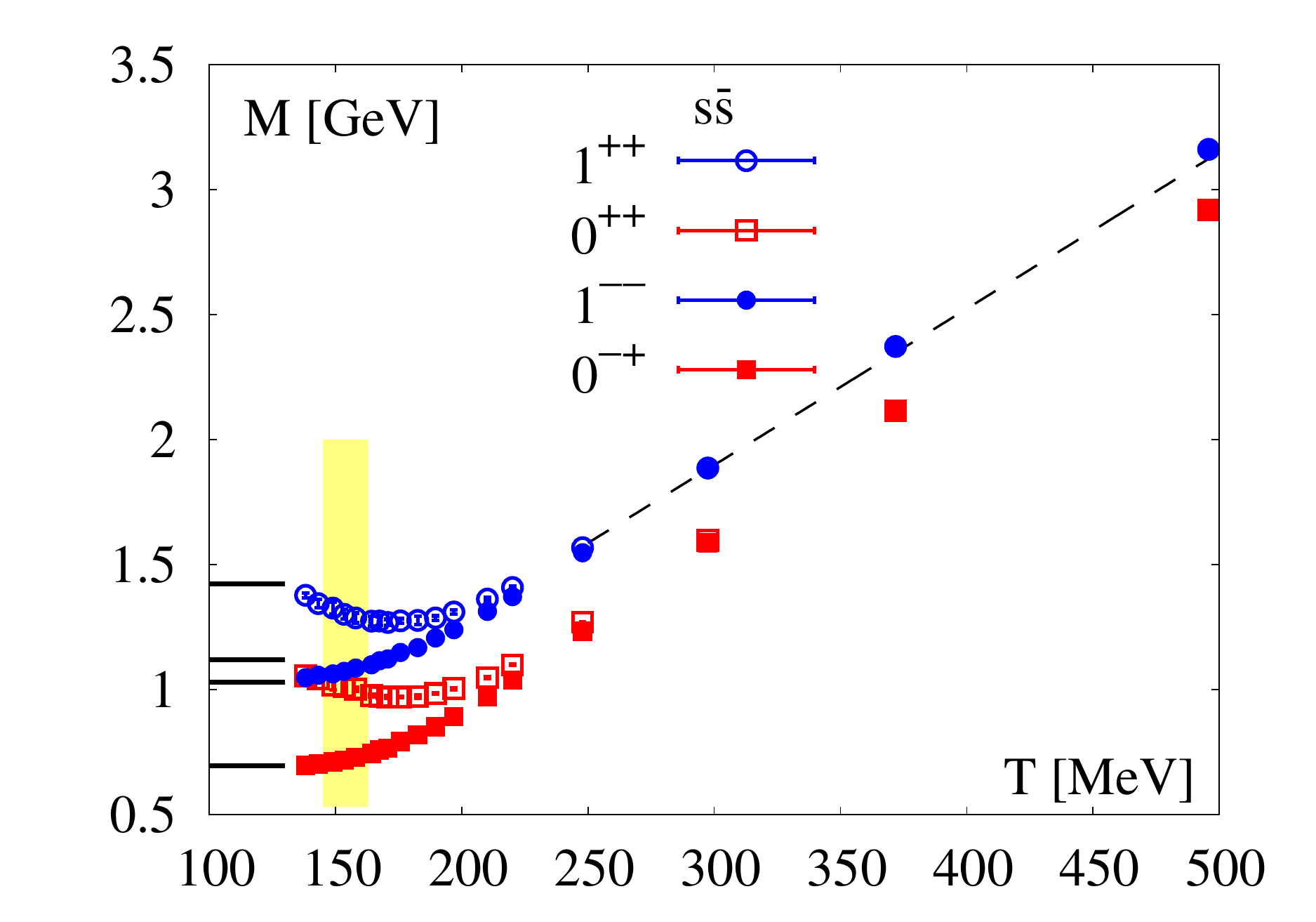}
 \hspace*{-0.5cm}
 \includegraphics[width=0.35\textwidth]{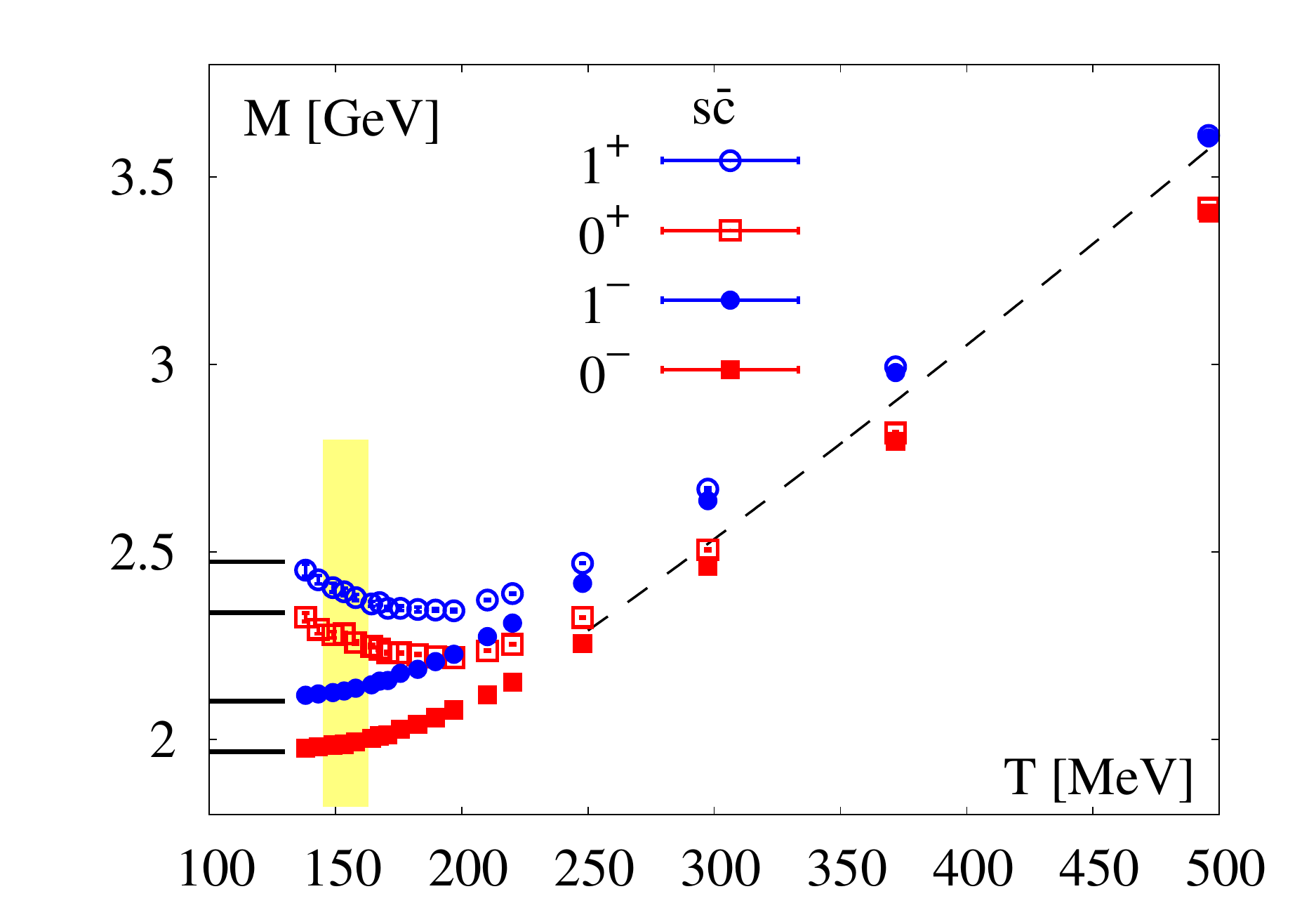}
 \hspace*{-0.5cm}
 \includegraphics[width=0.35\textwidth]{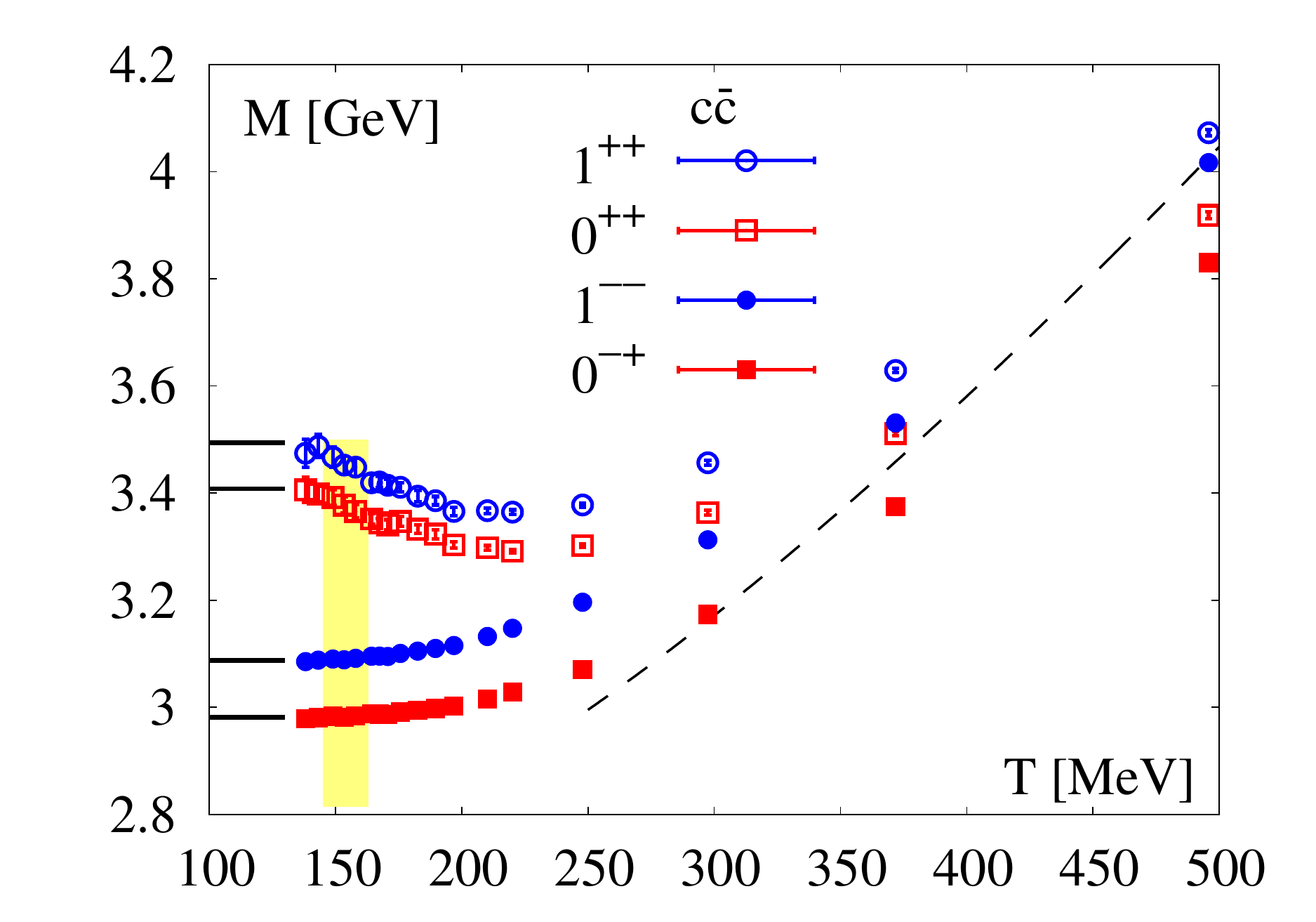}
 \caption{Meson screening masses for different channels in $s\bar s$ (left), $s\bar c$
(middle) and $c \bar c$ (right) sectors. The solid
horizontal lines on the left depict the corresponding zero temperature meson masses,
the shaded regions indicate the chiral crossover temperature $T_c=(154\pm9)$ MeV and
the dashed lines are the corresponding free field theory result. Taken from \cite{Bazavov:2014cta}.}
  }
  \label{fig:charm2}
\end{figure}

The analysis of generalized susceptibilities is limited to the open charm sector as flavor singlet meson states, like charmonium states, do not contribute here. Spatial correlation functions and screening masses defined by the exponential decay of the correlation functions at large distances are suitable observables to study the in-medium modifications of hadrons and the dissociation of individual states. Although they do not allow for a direct determination of spectral properties of these states (see next chapter for this), they have the advantage that they can be calculated rather easily already on rather small lattices. Results for screening masses of strange mesons, open strange-charm mesons and charmonium from \cite{Bazavov:2014cta} are shown in Fig.~\ref{fig:charm2}. 
Although medium modifications of the
spatial meson correlation functions set in close to the crossover
region, the amount of in-medium modifications
in the spatial correlators is different in different
sectors and decreases with the heavy quark content.
This is in agreement with a sequential melting of different states. Especially in the charmonium sector a shift of thermal modifications to higher temperatures compared to the lighter states including strange quarks is visible as well as a sequential melting of different charmonia states.
Larger and more loosely bound P-wave states may dissociate at lower temperatures
than the smaller and tightly bound S-wave charmonia.
For a more detailed description of in-medium modifications of hadrons a determination of the corresponding spectral functions from temporal correlation functions would be very useful in the future. Unfortunately this requires very large and fine lattices which so far where only available in the quenched approximation. Methods and results on this will be discussed in the following chapter and will provide the methodology to tackle these questions also in full QCD caluclations in the future.

\subsection{Conclusions}
We have presented recent progress in the studies of lattice QCD thermodynamics, with particular emphasis on different aspects involving strange quarks.
The discussion is based on a Taylor expansion of the QCD thermodynamical potential with respect to chemical potentials for baryon number, electric charge and strangeness. The so-defined generalized susceptibilities 
allow for a direct comparison with the phenomenology of the confined phase of QCD, represented by the hadron resonance gas model on the one hand and ratios of identified particles produced in heavy-ion collisions.
Appropriate combinations of conserved net strange and net charm fluctuations and their correlations with other conserved charges provide evidence that in the hadronic phase so far unobserved hadrons contribute to the thermodynamics and need to be included in hadron resonance gas models. 
In the strange sector this leads to significant reductions of the chemical freeze-out temperature of strange hadrons. 
We have presented a discussion of data from heavy-ion collisions at SPS, RHIC\index{RHIC} and LHC\index{LHC} in terms of the sudden chemical freeze-out model. 
It was found that a description of the thermodynamics of open strange and open charm degrees of freedom in terms of an uncorrelated hadron gas is valid only up to temperatures close to the chiral crossover\index{chiral!crossover} temperature. 
This suggests that in addition to light and strange hadrons also open charm hadrons start to dissolve already close to the chiral crossover\index{chiral!crossover}. 

In the remainder of this chapter we discuss some basics of lattice gauge theory and Monte Carlo calculations. This will provide the required knowledge for performing first lattice calculations for SU(3) pure gauge theory and studying thermodynamic quantities in the exercises of this chapter.  
Recent progress in extracting spectral and transport properties from lattice QCD calculations will be addressed 
separately in the following chapter.

\section{Basics of lattice gauge theory}

Lattice QCD is a framework in which the theory of strong interactions can be studied from first principles. It is based on the formulation of the theory of QCD in Euclidean space and discretizing space and Euclidean time. The Euclidean path integral allows to define the QCD partition function 
that controls equilibrium thermodynamics and is the basis for the calculation of all thermodynamic observables of QCD thermodynamics. In this basic introduction we will only focus on gluonic part sector\index{gluon!sector} of QCD as a discussion of the discretization of fermionic degrees of freedom goes beyond the scope of this book. Nevertheless, the basic concepts of lattice field theory can already be well discusssed for gauge theories. The exercises at the end of the chapter will allow practicing the discussed concepts using a simple framework for performing Monte Carlo calculations of the SU(3) lattice gauge theory.

\subsection{discretization of space-time points}
The grand canonical partition function of QCD at temperature $T$, volume $V$ and vanishing chemical potentials for the different quark flavors can be written as a path integral over the gauge fields, $A_\mu$ and fermion fields $\psi_f$ of the different flavors $f$ as
\begin{eqnarray}
{\cal{Z}}(T,V)=\int\prod_{\mu}{\cal D} A_{\mu}
\prod_{f}
{\cal D}\psi_f{\cal D}\bar{\psi}_f\ {\rm e}^{-S_E(T,V)} \; ,
\label{eq:partitionfunction}
\end{eqnarray}
with the Euclidean action
\begin{eqnarray}
S_E(T,V) =  - \int\limits_0^{1/T}
{\cal D} x_0 \int\limits_V {\cal D}^3 {\bf x} \;
{\cal L}^{E}_{QCD} (A_\mu,\psi_f,\bar\psi_f) \;,
\end{eqnarray}
which in continuous space-time is an integral of the QCD Lagrangian, ${\cal L}^{E}_{QCD}$, over space and time. The path integral defined in this way needs to be regularized. 

In lattice gauge theory, as suggested by K.G. Wilson in 1974 \cite{Wilson:1974sk},
space and time is discretized on a  
hyper-cubic lattice of size $N_{\sigma}^3\times N_{\tau}$
by introducing a finite lattice spacing $a$ (cut-off).
This is a particular regularization scheme of the path integral by introducing a finite ultra-violet cut-off. This leads to high dimensional integrals over the
gauge and fermion field variables that renders the path integral finite. 

The temperature $T$ and spatial volume $V$ is then given by
\begin{align*}
  T &= 1/(N_\tau \, a) \;,\\
  V &= L^3 =  (N_\sigma \, a)^3 \;.
\end{align*} 

The lattice spacing, $a$, serves as a cut-off which regularizes the ultraviolet divergences of the quantum field theory. In most lattice calculations an isotropic lattice is introduced, but sometimes also anisotropic lattices with different lattice spacings in temporal and spatial directions are used.
The discretization leads to systematic cut-off effects in obserables calculated on finite lattices. Therefore the lattice spacing needs to be taken to zero at the end of the calculations to obtain continuum results. This corresponds to taking the limit of
$N_\tau \rightarrow \infty$ at fixed temperature $T$, while appropriately tuning up $N_\sigma$ to keep the volume finite and large or taking the limit $N_\sigma \rightarrow \infty$, i.e. also performing the thermodynamic limit.
A large enough volume means that it sufficiently encompasses the various physical length scales in the problem. One of the longest scale is the pion wavelength and hence most lattice QCD calculations require $m_\pi L \gg 1$ to reduce any finite volume effects. In finite temperature calculations this usually leads to temporal extents, $N_t$, which are much smaller than the spatial extents, $N_\sigma$. Typical aspect ratios, $N_\sigma/N_t$, are of the order of 4 for physical pion masses but need to be increased for smaller than physical pion masses, e.g. in studies of the chiral limit\index{chiral!limit} of QCD. 

Although the discretized version of the partition function is finite, still a direct calculation of ${\cal Z}$ is usually not possible. Rather than calculating it directly,  
thermal expectation value of physical observables ${\cal O}$ can be obtained 
through
\begin{eqnarray}
\left<{\cal O}\right> &=& \frac{1}{Z(T,V)} 
\int\prod_{\mu}{\cal D} A_{\mu}\prod_{f}{\cal D}\psi_f{\cal D}\bar{\psi}_f\, {\cal O}\, 
{\rm e}^{-S_E(T,V)}  \;,
\label{eq:thermal_average}
\end{eqnarray}
using Monte-Carlo techniques as will be described in section \ref{sec:MonteCarlo}.

\subsection{Gauge transformation and gauge action}

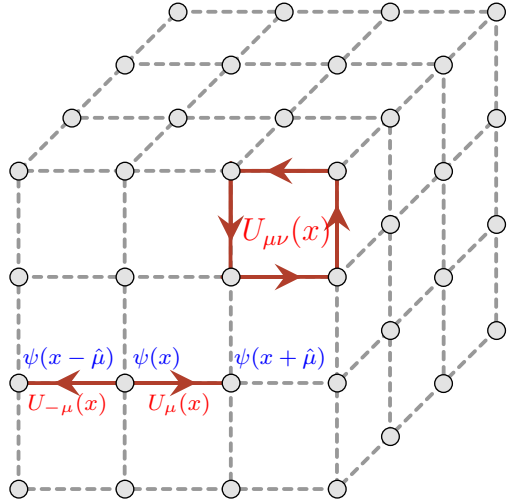
\begin{figure}[htb]
\begin{center}
\begin{tikzpicture}[      
       every node/.style={anchor=north west,inner sep=0pt}
      ]   
     \node (fig1) at (0,0)
         {\includegraphics[scale=0.5]{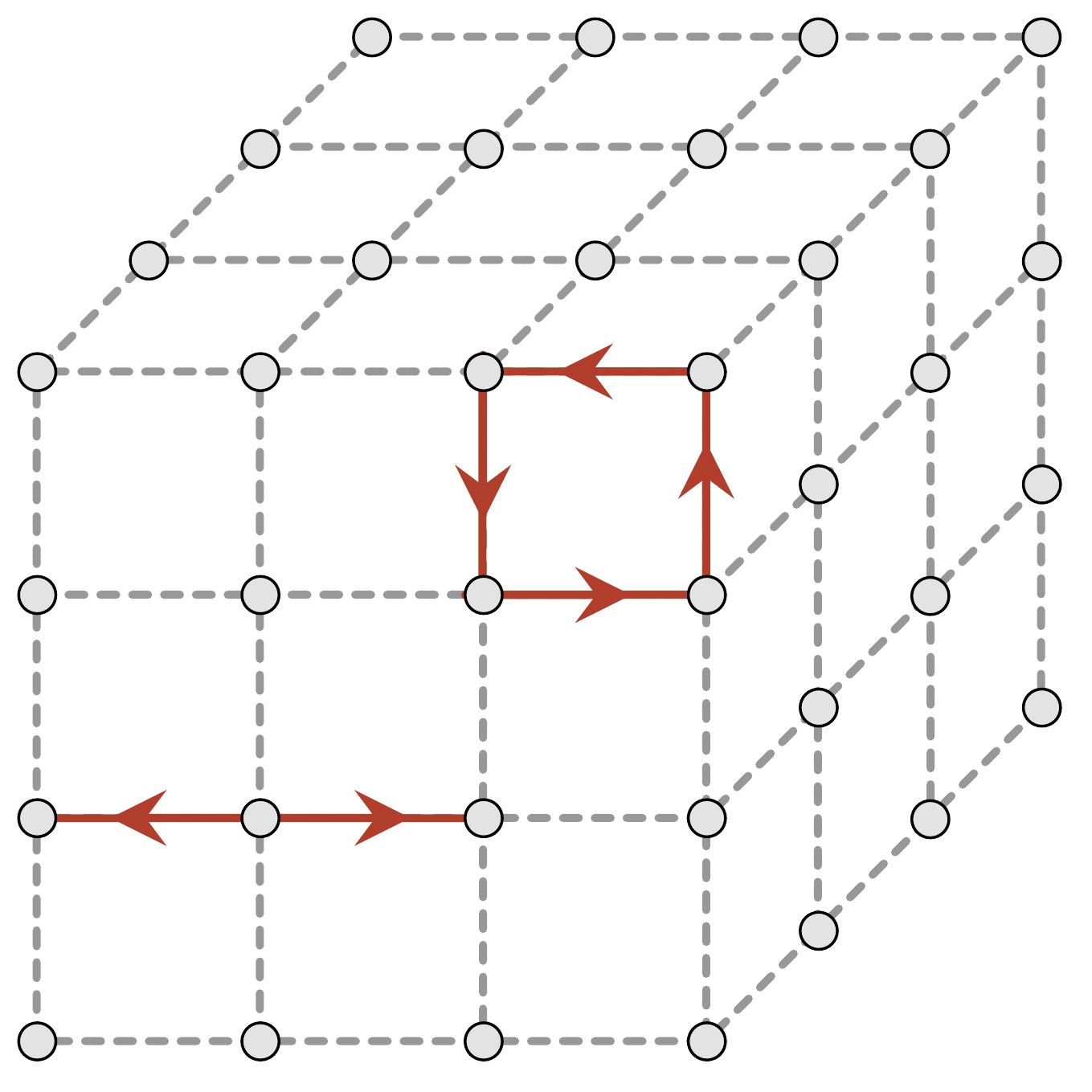}};
     \node (real) at (3.2,-2.95)
     {\textcolor{red}{\large$U_{\mu\nu}(x)$}};
     \node (imag) at (1.75,-4.7)
    {\textcolor{blue}{\small$\psi(x)$}};
     \node (imag) at (3.1,-4.7)
    {\textcolor{blue}{\small$\psi(x+\hat{\mu})$}};
     \node (imag) at (0.29,-4.7)
    {\textcolor{blue}{\small$\psi(x-\hat{\mu})$}};
     \node (imag) at (1.95,-5.3)
    {\textcolor{red}{\small$U_{\mu}(x)$}};
     \node (imag) at (0.35,-5.3)
    {\textcolor{red}{\small$U_{-\mu}(x)$}};
\end{tikzpicture}
  \caption{Lattice implementation of QCD: fermion fields $\psi(x)$ (blue) living on the sites and gauge link variables $U_\mu(x)$ living on the links. The plaquette $U_{\mu\nu}(x)$ is a product of links enclosing the smallest closed path. Picture by Lukas Mazur, Bielefeld University.}
\label{fig:plaquette}
\end{center}
\end{figure}

In the lattice approach to QCD, the quark field $\psi(x)$ and $\bar \psi(x)$ are defined on the sites of the lattice as anticommuting Grassmann variables, 
while the gauge fields $A^a_\mu(x), a = 1, 2, ..., 8$ 
are defined on the links connecting the nearest neighboring sites via the link variable $U_\mu(x)$,
\begin{align}
 U_\mu(x)  = \mathrm{P} \ \mathrm{exp} \left( i g \int_x^{x+a\hat\mu} \mathrm{d}x_\mu \, T^a A^a_\mu(x)\right) \;,
\end{align}
where P denotes path ordering.
The gauge links live on the site connecting $x$ and $x+a\hat\mu$
where $\hat{\mu}$ is a unit vector along $\mu$-th direction and $T^a$ are the eight generators of the SU($N_c=3$) color group\index{color!SU(3) group}.
Under local gauge transformation $\Omega(x) = \exp \, [ -i \alpha^a(x) \, T^a]$, the fields $U_\mu(x)$ and $\psi(x)$ transform as
\begin{align}
  \begin{split}
  \psi(x) &\rightarrow \Omega(x) \psi(x) \\
   \bar{\psi}(x) &\,\rightarrow\, \bar{\psi}(x) \Omega(x)^\dag \\
    U_\mu(x)  &\rightarrow \Omega(x + a\,\hat{\mu}) U_\mu(x) \Omega^{\dag} (x).
  \end{split}
\label{eq:gauge_transform}
\end{align}
This distribution of the fermion and gauge fields allows for a gauge invariant discretization of the QCD action on the lattice. 
The simplest gauge invariant object on the lattice is the trace of the product of links enclosing the smallest closed path, i.e.
the plaquette $U_{\mu\nu}(x)$ (see Fig. \ref{fig:plaquette}):

\begin{equation}
  U_{\mu\nu}(x) = U_\mu(x) U_\nu(x+\hat\mu) U^\dagger_\mu(x+\hat\nu) U^\dagger_\nu(x) \;.
\end{equation} 
%
%


\noindent 

The gauge invariant trace of the plaquettes can be used to formulate the simplest version of the discretization of the gluonic action\index{gluon!action}, the so called Wilson gauge action,
\begin{align}
  S_G &= \beta \sum_{x} \sum_{\mu > \nu} \, (1-\frac{1}{N_c} \, {\rm Re} \, {\rm Tr}_c \, U_{\mu\nu}(x)) \;,
  \label{eq:gauge_action}
\end{align} 
with $\beta=\frac{2 N_c}{g^2}$.
In the continuum limit, $a\rightarrow 0$, this action reproduces the continuum gauge action and at non-zero lattice spacing has discretization errors of ${\cal O}(a^2)$. 
These systematic errors introduced by the finite lattice cut-off can be reduced using improved discretization schemes.

A detailed discussion of the discretization of the fermionic part of the QCD action goes beyond the scope of this book and can be found in the textbooks \cite{Gattringer:2010zz,Montvay:1994cy,Rothe:1992nt,DeGrand:2006zz}.

\subsection{Renormalization and continuum limit}

In the gauge action~\eqref{eq:gauge_action} we introduce a bare parameter, the coupling constant $g$. To ensure a correct continuum limit, $a\rightarrow 0$, and physical results in the continuum that do not depend on the lattice spacing and the specific discretization of the theory, the coupling and also other bare parameters, like quark masses, have a non-trivial dependence on the lattice spacing. Or turning this around, the lattice spacing $a$ depends on the bare parameters. Therefore, in the pure gauge theory, the bare coupling g is the only parameter that provides a handle to tune the lattice spacing $a$.
The dependence of the lattice spacing on the bare coupling, correspondingly the running of the coupling with the lattice spacing is controlled by the renormalization group. For small values of the bare gauge coupling this is controlled
by the QCD $\beta$-function,
\begin{eqnarray}
a \Lambda_L = \left( \frac{1}{b_0g^2}\right)^{b_1/2b_0^2} {\rm e}^{-1/2b_0g^2} \;  (1+{\cal O}(g^2)) \;, 
\label{betafunction}
\end{eqnarray}
with $b_0=(\frac{11}{3}N_c-\frac{2}{3}N_f)/16\pi^2$ and
$b_1=( \frac{34}{3}N_c^2 -( \frac{10}{3}N_c + \frac{N_c^2-1}{N_c} 
)N_f)/(16\pi^2)^2$. The parameters $b_0$ and $b_1$ are universal and do not dependent on the specific regularization of the theory, while the higher order terms are in general non-universal. The limit of vanishing lattice spacing $a$ corresponds to the limit of vanishing gauge coupling $g$, i.e. asymptotic freedom\index{asymptotic freedom}. As the QCD $\beta$-function has a UV fixed point\index{fixed point} at $g=0$ implies that with a proper tuning of the bare parameters, on a line of constant physics, physical observables can be extracted using lattice calculations at finite lattice spacing $a$ and that a non-trivial continuum limit exists.

Although the pure gauge theory has no dimensional parameter, the renormalization introduces a mass scale, $\Lambda_L$, and all physical quantities with a dimension are proportional to appropriate powers of $\Lambda_L$, which is the integration constant in (\ref{betafunction}). This allows to extract physical observables, up to remaining cut-off effects, already on finite lattices and allows to set the scale, i.e. determining the lattice spacing $a$ in physical units, from calculating observables on the lattice and matching these to their physical values obtained from experiment. 

\section{Basics of Monte Carlo integration}\label{sec:MonteCarlo}

Although discretizing space and time changes the infinite dimensional integrations appearing in the partition function Eq.~(\ref{eq:partitionfunction}) and in the calculation of observables Eq.~(\ref{eq:thermal_average}) to high dimensional integrations, in most situations it is impossible to do these integrations analytically or using standard numerical integration methods.
The basic idea to the Monte Carlo integration technique to studying statistical systems\index{statistical!systems}
is to obtain a suitable distribution of states.
Given that one can determine the value of the Hamiltonian for a given state, 
the task is then to construct an algorithm to obtain an ensemble in which configurations appear according to the Boltzmann\index{Boltzmann!statistics} weight.
In this section we denote the fields as $\phi$, which in the pure gauge theory can be interpreted as the gauge fields $U_{x,\mu}$, but the discussion is general for arbitrary fields.

\subsection{Importance sampling}
\label{Importance_sampling}
The expectation value of a physical
observable $O$ is represented as
\begin{eqnarray}
  \langle O \rangle &=&  \frac{1}{Z} \int \mathcal{D}[\phi]
   \ O[\phi] \ e^{-S[\phi]}\\
   &=& \int \mathcal{D}[\phi]
   \ O[\phi] W[\phi],\label{eq:boltzmann_weight}
\end{eqnarray}
where $\phi$ denotes collectively the degrees of freedom and $Z$ on the right hand is the partition function defined as 
\begin{align}
  Z=\int \mathcal{D}[\phi]\  \text{e}^{-S_E[\phi]},
\end{align}
where $S_E$ is the Euclidean action. For example, for the pure gauge theory the Wilson gauge action Eq.~(\ref{eq:gauge_action}) can used here. In that case the functional integral in the above equations turns into high dimensional integrals over all gauge fields $U_{x,\mu}$.
In the case of real actions, the Boltzmann weight,
\begin{eqnarray}
W[\phi]=\frac{1}{Z} e^{-S[\phi]}\;, 
\label{eq:Boltzmann_weight2}
\end{eqnarray}
appearing in Eq.~(\ref{eq:boltzmann_weight}) can be interpreted as a probability density, i.e.
\begin{eqnarray}
W[\phi] \geq 0 \ \; , \ \int\mathcal{D}[\phi] W[\phi] = 1.
\end{eqnarray}
If this probability density is available, there is no need to calculate the partition function to calculate physical observables. The idea of Monte Carlo calculations is to sample the states of the system, i.e. the field configurations, using importance sampling, where the configurations are sampled with a distribution according to the corresponding probability density. Generating a finite number, $N$, of field configurations, $\{\phi\}_l$, $l=1,\ldots N$, the integral in (\ref{eq:boltzmann_weight}) can be calculated statistically
by an average over these specific configurations,
\begin{align}
  \langle O \rangle \approx \frac{1}{N}
     \sum_i O[\phi_i] + \mathcal{O}(\frac{1}{\sqrt{N}}) .
\end{align}
This statistical integration\index{statistical!integration}, based on field configurations that are weighted according to the Boltzmann weight, is called importance sampling and allows to calculate expectation values of $O$ and its statistical errors\index{statistical!error} using a finite number of field configurations. As the probability density is usually not known, methods are required to generate field configurations respecting this. Most algorithms are based on the concept of Markov chains which will be discussed in the following.

\subsection{Markov chain}

The idea of Markov chains is to generate a sequence of field configurations $\{\phi\}_l$ using a sequential stochastic process,
\begin{eqnarray}
\{\phi\}_0 \rightarrow \{\phi\}_1 \rightarrow \{\phi\}_2 \rightarrow \ldots
\end{eqnarray}
starting with an initial configuration $\{\phi\}_0$ and such that, after some steps of thermalization\index{thermalization} of the process, the probability to find a configuration in $\lbrack \phi, \phi+\delta\phi\rbrack$ is given by $w[\phi]\delta\phi$.

The idea is to construct a stochastic process with a   
transition probability,
\begin{eqnarray}
P (\phi \rightarrow \phi') \ \ \textrm{with} \ \ \sum_{\phi'} P (\phi \rightarrow \phi') = 1,
\end{eqnarray}
for the subsequent steps in the Markov chain. Using the concept of ergodicity,
\begin{eqnarray}
P (\phi \rightarrow \phi') > 0 \ \ \ \forall \phi,\phi'\;,
\end{eqnarray}
which ensures a finite probability to reach any state $\phi'$ from any other state $\phi$, one can proof the existence of a unique fixed point\index{fixed point} of the stochastic process and furthermore that the distribution of states converges to the fixed fixed point, independent of the start configuration.
A sufficient (not necessary) condition for the process to reach the desired equilibrium distribution given by the Boltzmann weight Eq.~(\ref{eq:Boltzmann_weight2})
is the detailed balance condition,
\begin{align}
  e^{-S[\phi]} \, P[\phi \rightarrow \phi^\prime] = e^{-S[\phi^\prime]} \, P[\phi^\prime \rightarrow \phi].
\end{align}
Therefore most algorithms are based on this condition, e.g. the Metropolis algorithm \cite{Metropolis:1953am} developed in the year 1953 is a prototype of a Monte Carlo algorithm and applicable to any statistical system\index{statistical!system}.

In pure gauge theories it is common to combine two algorithms. In the heat-bath algorithm single gauge links, $U_{x,\mu}$, are updated by choosing a new link $U'_{x,\mu}$ that depends only on the interaction with its immediate surrounding links, while the other links are treated as the heat-bath. The probability for updating is proportional to
\begin{align}
  P(U) \propto e^{-\frac{\beta}{N_c} \, {\rm Re \, Tr} U A}~,
\end{align}
where $A$ represents the staple, i.e. the local parts of the action that are connected to but do not contain $U_{x,\mu}$ 
This local update process is iterated for all links of the lattice (a sweep), eventually covering the whole system, providing a new gauge field configuration.
For SU(2) gauge theory the heat-bath algorithm was found by Creutz \cite{Creutz:1980zw}
and for SU(3) a pseudo heat-bath algorithm developed by Cabibbo and Marinari \cite{Cabibbo:1982zn} is based on updating SU(2) subgroups of the SU(3) links. 

Generated gauge field configurations using such local algorithms are highly correlated and it takes a large number of sweeps to obtain statistically independent configurations. Therefore the heat-bath algorithm is usually combined with an overrelaxation update \cite{Adler:1981sn} to reduce auto-correlations.
The idea is to find a new link $U'_{x,\mu}$ which is \textit{far away} from $U_{x,\mu}$, but has the same probability weight. Therefore this new link doesn't change the action and the algorithm alone is not ergodic, but in combination with the heat-bath algorithm this leads to a large reduction of auto-correlations. 
For SU(2), SU(3) and general SU(N) gauge theories discussions on overrelaxation algorithms can be found in \cite{Brown:1987rra,Gupta:1984gq,Creutz:1987xi}.

\section{Exercises}
In the exercises for this chapter we use a simple Monte Carlo program for SU(3) pure gauge theory with Wilson gauge action written in C++. 

The program can be download from\medskip\\
\centerline{\href{https://www.physik.uni-bielefeld.de/~okacz/su3.tgz}{https://www.physik.uni-bielefeld.de/$\sim$okacz/su3.tgz}}\medskip\\
and compiled using \textit{make} resulting in an executable \textit{su3\_run} which can be run using \textit{su3\_run in.sample}, where \textit{in.sample} is a parameter file containing parameters like the lattice size, beta values, file names for the result files and others.
The program uses a combination of the heatbath and overrelaxation algorithms and calculates as observables the plaquette and Polyakov loop\index{Polyakov loop}.\\\\

\noindent Exercise 1\\\\
Take a look into the main program in \textit{su3\_run.cpp} and the parameter file \textit{in.sample} and try to understand the usage and flow of the program, especially how the gauge fields are defined and used and how the updates and measurements of the observables are programmed.\\\\

\noindent Exercise 2\\\\
Take a look into the paper "Critical point\index{critical!point} and scale setting in SU(3) plasma: An
update" \cite{Francis:2015lha}. You will find the critical values for
various $N_t$ and typical lattice volumes and the scale for the Wilson gauge action.
Estimate the run-time of the program on your computer to generate measurements with \cal{O}(10.000) sweeps. Choose an appropriate lattice size and perform calculations at some temperatures around the critical temperature. Study the thermalization behavior for different start configurations (random vs. unit) and study the behavior of the Polyakov loop, the order parameter for the deconfinement\index{deconfinement} transition in the SU(3) gauge theory (after thermalization of the Monte Carlo sweeps).\\\\

\noindent Exercise 3\\\\
Take a look into the famous Boyd et al. paper \cite{Boyd:1996bx} on the 
"Thermodynamics of SU(3) Lattice Gauge Theory" and reproduce Fig.~3 of that paper. If your computer is fast enough, you may add a finer lattice, e.g. for $N_t=10$.
Follow the discussion on the derivation of thermodynamic observables and use your results together with the scale setting of \cite{Francis:2015lha} to reproduce the results for the pressure and energy density.


%
%




\begin{thebibliography}{10}

\bibitem{Gattringer:2010zz}
C.~Gattringer and C.~B. Lang,
  \href{http://dx.doi.org/10.1007/978-3-642-01850-3}{{\em {Quantum
  chromodynamics on the lattice}}}, vol.~788.
\newblock Springer, Berlin, 2010.

\bibitem{Montvay:1994cy}
I.~Montvay and G.~Munster,
  \href{http://dx.doi.org/10.1017/CBO9780511470783}{{\em {Quantum fields on a
  lattice}}}.
\newblock Cambridge Monographs on Mathematical Physics. Cambridge University
  Press, 3, 1997.

\bibitem{Rothe:1992nt}
H.~J. Rothe, {\em {Lattice gauge theories: An Introduction}}, vol.~82.
\newblock World Sci.Lect.Notes Phys., 2012.

\bibitem{DeGrand:2006zz}
T.~DeGrand and C.~E. Detar, {\em {Lattice methods for quantum chromodynamics}}.
\newblock World Scientific, 2006.

\bibitem{Karsch:2003jg}
F.~Karsch and E.~Laermann, ``{Thermodynamics and in medium hadron properties
  from lattice QCD},'' \href{http://arxiv.org/abs/hep-lat/0305025}{{\tt
  arXiv:hep-lat/0305025}}.

\bibitem{Ding:2015ona}
H.-T. Ding, F.~Karsch, and S.~Mukherjee, ``{Thermodynamics of
  strong-interaction matter from Lattice QCD},''
  \href{http://dx.doi.org/10.1142/S0218301315300076}{{\em Int. J. Mod. Phys. E}
  {\bf 24} (2015) no.~10, 1530007}, \href{http://arxiv.org/abs/1504.05274}{{\tt
  arXiv:1504.05274 [hep-lat]}}.

\bibitem{Guenther:2020jwe}
J.~N.~Guenther,
Eur. Phys. J. A \textbf{57}, no.4, 136 (2021)
doi:10.1140/epja/s10050-021-00354-6
[arXiv:2010.15503 [hep-lat]].

\bibitem{Rothkopf:2019ipj}
A.~Rothkopf, ``{Heavy Quarkonium in Extreme Conditions},''
  \href{http://dx.doi.org/10.1016/j.physrep.2020.02.006}{{\em Phys. Rept.} {\bf
  858} (2020)  1--117}, \href{http://arxiv.org/abs/1912.02253}{{\tt
  arXiv:1912.02253 [hep-ph]}}.

\bibitem{Bazavov:2011nk}
A.~Bazavov {\em et al.}, ``{The chiral and deconfinement aspects of the QCD
  transition},'' \href{http://dx.doi.org/10.1103/PhysRevD.85.054503}{{\em Phys.
  Rev. D} {\bf 85} (2012)  054503}, \href{http://arxiv.org/abs/1111.1710}{{\tt
  arXiv:1111.1710 [hep-lat]}}.

\bibitem{Bhattacharya:2014ara}
T.~Bhattacharya {\em et al.}, ``{QCD Phase Transition with Chiral Quarks and
  Physical Quark Masses},''
  \href{http://dx.doi.org/10.1103/PhysRevLett.113.082001}{{\em Phys. Rev.
  Lett.} {\bf 113} (2014) no.~8, 082001},
  \href{http://arxiv.org/abs/1402.5175}{{\tt arXiv:1402.5175 [hep-lat]}}.

\bibitem{Aoki:2006we}
Y.~Aoki, G.~Endrodi, Z.~Fodor, S.~D. Katz, and K.~K. Szabo, ``{The Order of the
  quantum chromodynamics transition predicted by the standard model of particle
  physics},'' \href{http://dx.doi.org/10.1038/nature05120}{{\em Nature} {\bf
  443} (2006)  675--678}, \href{http://arxiv.org/abs/hep-lat/0611014}{{\tt
  arXiv:hep-lat/0611014}}.

\bibitem{Ding:2019prx}
H.~T. Ding {\em et al.}, ``{Chiral Phase Transition Temperature in ( 2+1
  )-Flavor QCD},'' \href{http://dx.doi.org/10.1103/PhysRevLett.123.062002}{{\em
  Phys. Rev. Lett.} {\bf 123} (2019) no.~6, 062002},
  \href{http://arxiv.org/abs/1903.04801}{{\tt arXiv:1903.04801 [hep-lat]}}.

\bibitem{Bazavov:2014pvz}
{\bf HotQCD} Collaboration, A.~Bazavov {\em et al.}, ``{Equation of state in (
  2+1 )-flavor QCD},'' \href{http://dx.doi.org/10.1103/PhysRevD.90.094503}{{\em
  Phys. Rev. D} {\bf 90} (2014)  094503},
  \href{http://arxiv.org/abs/1407.6387}{{\tt arXiv:1407.6387 [hep-lat]}}.

\bibitem{Borsanyi:2013bia}
S.~Borsanyi, Z.~Fodor, C.~Hoelbling, S.~D. Katz, S.~Krieg, and K.~K. Szabo,
  ``{Full result for the QCD equation of state with 2+1 flavors},''
  \href{http://dx.doi.org/10.1016/j.physletb.2014.01.007}{{\em Phys. Lett. B}
  {\bf 730} (2014)  99--104}, \href{http://arxiv.org/abs/1309.5258}{{\tt
  arXiv:1309.5258 [hep-lat]}}.

\bibitem{Bazavov:2018mes}
{\bf HotQCD} Collaboration, A.~Bazavov {\em et al.}, ``{Chiral crossover in QCD
  at zero and non-zero chemical potentials},''
  \href{http://dx.doi.org/10.1016/j.physletb.2019.05.013}{{\em Phys. Lett. B}
  {\bf 795} (2019)  15--21}, \href{http://arxiv.org/abs/1812.08235}{{\tt
  arXiv:1812.08235 [hep-lat]}}.

\bibitem{Cheng:2007jq}
M.~Cheng {\em et al.}, ``{The QCD equation of state with almost physical quark
  masses},'' \href{http://dx.doi.org/10.1103/PhysRevD.77.014511}{{\em Phys.
  Rev. D} {\bf 77} (2008)  014511}, \href{http://arxiv.org/abs/0710.0354}{{\tt
  arXiv:0710.0354 [hep-lat]}}.

\bibitem{Karsch:2001cy}
F.~Karsch, ``{Lattice QCD at high temperature and density},''
  \href{http://dx.doi.org/10.1007/3-540-45792-5_6}{{\em Lect. Notes Phys.} {\bf
  583} (2002)  209--249}, \href{http://arxiv.org/abs/hep-lat/0106019}{{\tt
  arXiv:hep-lat/0106019}}.

\bibitem{Zyla:2020zbs}
{\bf Particle Data Group} Collaboration, P.~A. Zyla {\em et al.}, ``{Review of
  Particle Physics},'' \href{http://dx.doi.org/10.1093/ptep/ptaa104}{{\em PTEP}
  {\bf 2020} (2020) no.~8, 083C01}.

\bibitem{Capstick:1986bm}
S.~Capstick and N.~Isgur, ``{Baryons in a Relativized Quark Model with
  Chromodynamics},'' \href{http://dx.doi.org/10.1103/PhysRevD.34.2809}{{\em AIP
  Conf. Proc.} {\bf 132} (1985)  267--271}.

\bibitem{Ebert:2009ub}
D.~Ebert, R.~N. Faustov, and V.~O. Galkin, ``{Mass spectra and Regge
  trajectories of light mesons in the relativistic quark model},''
  \href{http://dx.doi.org/10.1103/PhysRevD.79.114029}{{\em Phys. Rev. D} {\bf
  79} (2009)  114029}, \href{http://arxiv.org/abs/0903.5183}{{\tt
  arXiv:0903.5183 [hep-ph]}}.

\bibitem{Edwards:2012fx}
{\bf Hadron Spectrum} Collaboration, R.~G. Edwards, N.~Mathur, D.~G. Richards,
  and S.~J. Wallace, ``{Flavor structure of the excited baryon spectra from
  lattice QCD},'' \href{http://dx.doi.org/10.1103/PhysRevD.87.054506}{{\em
  Phys. Rev. D} {\bf 87} (2013) no.~5, 054506},
  \href{http://arxiv.org/abs/1212.5236}{{\tt arXiv:1212.5236 [hep-ph]}}.

\bibitem{Bazavov:2014xya}
A.~Bazavov {\em et al.}, ``{Additional Strange Hadrons from QCD Thermodynamics
  and Strangeness Freezeout in Heavy Ion Collisions},''
  \href{http://dx.doi.org/10.1103/PhysRevLett.113.072001}{{\em Phys. Rev.
  Lett.} {\bf 113} (2014) no.~7, 072001},
  \href{http://arxiv.org/abs/1404.6511}{{\tt arXiv:1404.6511 [hep-lat]}}.

\bibitem{Andronic:2017pug}
A.~Andronic, P.~Braun-Munzinger, K.~Redlich, and J.~Stachel, ``{Decoding the
  phase structure of QCD via particle production at high energy},''
  \href{http://dx.doi.org/10.1038/s41586-018-0491-6}{{\em Nature} {\bf 561}
  (2018) no.~7723, 321--330}, \href{http://arxiv.org/abs/1710.09425}{{\tt
  arXiv:1710.09425 [nucl-th]}}.

\bibitem{NA57:2004nxc}
{\bf NA57} Collaboration, F.~Antinori {\em et al.}, ``{Energy dependence of
  hyperon production in nucleus nucleus collisions at SPS},''
  \href{http://dx.doi.org/10.1016/j.physletb.2004.05.025}{{\em Phys. Lett. B}
  {\bf 595} (2004)  68--74}, \href{http://arxiv.org/abs/nucl-ex/0403022}{{\tt
  arXiv:nucl-ex/0403022}}.

\bibitem{Zhao:2013yza}
{\bf STAR} Collaboration, F.~Zhao, ``{Beam Energy Dependence of Strange Hadron
  Production at RHIC},'' \href{http://dx.doi.org/10.22323/1.185.0036}{{\em PoS}
  {\bf CPOD2013} (2013)  036}.

\bibitem{Bazavov:2017dus}
A.~Bazavov {\em et al.}, ``{The QCD Equation of State to $\mathcal{O}(\mu_B^6)$
  from Lattice QCD},'' \href{http://dx.doi.org/10.1103/PhysRevD.95.054504}{{\em
  Phys. Rev. D} {\bf 95} (2017) no.~5, 054504},
  \href{http://arxiv.org/abs/1701.04325}{{\tt arXiv:1701.04325 [hep-lat]}}.

\bibitem{Gunther:2016vcp}
J.~N. Guenther, R.~Bellwied, S.~Borsanyi, Z.~Fodor, S.~D. Katz, A.~Pasztor,
  C.~Ratti, and K.~K. Szab\'o, ``{The QCD equation of state at finite density
  from analytical continuation},''
  \href{http://dx.doi.org/10.1016/j.nuclphysa.2017.05.044}{{\em Nucl. Phys. A}
  {\bf 967} (2017)  720--723}, \href{http://arxiv.org/abs/1607.02493}{{\tt
  arXiv:1607.02493 [hep-lat]}}.

\bibitem{Kaczmarek:2011zz}
O.~Kaczmarek, F.~Karsch, E.~Laermann, C.~Miao, S.~Mukherjee, P.~Petreczky,
  C.~Schmidt, W.~Soeldner, and W.~Unger, ``{Phase boundary for the chiral
  transition in (2+1) -flavor QCD at small values of the chemical potential},''
  \href{http://dx.doi.org/10.1103/PhysRevD.83.014504}{{\em Phys. Rev. D} {\bf
  83} (2011)  014504}, \href{http://arxiv.org/abs/1011.3130}{{\tt
  arXiv:1011.3130 [hep-lat]}}.

\bibitem{Endrodi:2011gv}
G.~Endrodi, Z.~Fodor, S.~D. Katz, and K.~K. Szabo, ``{The QCD phase diagram at
  nonzero quark density},''
  \href{http://dx.doi.org/10.1007/JHEP04(2011)001}{{\em JHEP} {\bf 04} (2011)
  001}, \href{http://arxiv.org/abs/1102.1356}{{\tt arXiv:1102.1356 [hep-lat]}}.

\bibitem{Bonati:2015bha}
C.~Bonati, M.~D'Elia, M.~Mariti, M.~Mesiti, F.~Negro, and F.~Sanfilippo,
  ``{Curvature of the chiral pseudocritical line in QCD: Continuum extrapolated
  results},'' \href{http://dx.doi.org/10.1103/PhysRevD.92.054503}{{\em Phys.
  Rev. D} {\bf 92} (2015) no.~5, 054503},
  \href{http://arxiv.org/abs/1507.03571}{{\tt arXiv:1507.03571 [hep-lat]}}.

\bibitem{Bellwied:2015rza}
R.~Bellwied, S.~Borsanyi, Z.~Fodor, J.~G\"unther, S.~D. Katz, C.~Ratti, and
  K.~K. Szabo, ``{The QCD phase diagram from analytic continuation},''
  \href{http://dx.doi.org/10.1016/j.physletb.2015.11.011}{{\em Phys. Lett. B}
  {\bf 751} (2015)  559--564}, \href{http://arxiv.org/abs/1507.07510}{{\tt
  arXiv:1507.07510 [hep-lat]}}.

\bibitem{Cea:2015cya}
P.~Cea, L.~Cosmai, and A.~Papa, ``{Critical line of 2+1 flavor QCD: Toward the
  continuum limit},'' \href{http://dx.doi.org/10.1103/PhysRevD.93.014507}{{\em
  Phys. Rev. D} {\bf 93} (2016) no.~1, 014507},
  \href{http://arxiv.org/abs/1508.07599}{{\tt arXiv:1508.07599 [hep-lat]}}.

\bibitem{Das:2014qca}
{\bf STAR} Collaboration, S.~Das, ``{Identified particle production and
  freeze-out properties in heavy-ion collisions at RHIC Beam Energy Scan
  program},'' \href{http://dx.doi.org/10.1051/epjconf/20159008007}{{\em EPJ Web
  Conf.} {\bf 90} (2015)  08007}, \href{http://arxiv.org/abs/1412.0499}{{\tt
  arXiv:1412.0499 [nucl-ex]}}.

\bibitem{Floris:2014pta}
M.~Floris, ``{Hadron yields and the phase diagram of strongly interacting
  matter},'' \href{http://dx.doi.org/10.1016/j.nuclphysa.2014.09.002}{{\em
  Nucl. Phys. A} {\bf 931} (2014)  103--112},
  \href{http://arxiv.org/abs/1408.6403}{{\tt arXiv:1408.6403 [nucl-ex]}}.

\bibitem{Becattini:2016xct}
F.~Becattini, J.~Steinheimer, R.~Stock, and M.~Bleicher, ``{Hadronization
  conditions in relativistic nuclear collisions and the QCD pseudo-critical
  line},'' \href{http://dx.doi.org/10.1016/j.physletb.2016.11.033}{{\em Phys.
  Lett. B} {\bf 764} (2017)  241--246},
  \href{http://arxiv.org/abs/1605.09694}{{\tt arXiv:1605.09694 [nucl-th]}}.

\bibitem{Bazavov:2014yba}
A.~Bazavov {\em et al.}, ``{The melting and abundance of open charm hadrons},''
  \href{http://dx.doi.org/10.1016/j.physletb.2014.08.034}{{\em Phys. Lett. B}
  {\bf 737} (2014)  210--215}, \href{http://arxiv.org/abs/1404.4043}{{\tt
  arXiv:1404.4043 [hep-lat]}}.

\bibitem{Bazavov:2014cta}
A.~Bazavov, F.~Karsch, Y.~Maezawa, S.~Mukherjee, and P.~Petreczky, ``{In-medium
  modifications of open and hidden strange-charm mesons from spatial
  correlation functions},''
  \href{http://dx.doi.org/10.1103/PhysRevD.91.054503}{{\em Phys. Rev. D} {\bf
  91} (2015) no.~5, 054503}, \href{http://arxiv.org/abs/1411.3018}{{\tt
  arXiv:1411.3018 [hep-lat]}}.

\bibitem{Wilson:1974sk}
K.~G. Wilson, ``{Confinement of Quarks},''
  \href{http://dx.doi.org/10.1103/PhysRevD.10.2445}{{\em Phys. Rev. D} {\bf 10}
  (1974)  2445--2459}.

\bibitem{Metropolis:1953am}
N.~Metropolis, A.~W. Rosenbluth, M.~N. Rosenbluth, A.~H. Teller, and E.~Teller,
  ``{Equation of state calculations by fast computing machines},''
  \href{http://dx.doi.org/10.1063/1.1699114}{{\em J. Chem. Phys.} {\bf 21}
  (1953)  1087--1092}.

\bibitem{Creutz:1980zw}
M.~Creutz, ``{Monte Carlo Study of Quantized SU(2) Gauge Theory},''
  \href{http://dx.doi.org/10.1103/PhysRevD.21.2308}{{\em Phys. Rev. D} {\bf 21}
  (1980)  2308--2315}.

\bibitem{Cabibbo:1982zn}
N.~Cabibbo and E.~Marinari, ``{A New Method for Updating SU(N) Matrices in
  Computer Simulations of Gauge Theories},''
  \href{http://dx.doi.org/10.1016/0370-2693(82)90696-7}{{\em Phys. Lett. B}
  {\bf 119} (1982)  387--390}.

\bibitem{Adler:1981sn}
S.~L. Adler, ``{An Overrelaxation Method for the Monte Carlo Evaluation of the
  Partition Function for Multiquadratic Actions},''
  \href{http://dx.doi.org/10.1103/PhysRevD.23.2901}{{\em Phys. Rev. D} {\bf 23}
  (1981)  2901}.

\bibitem{Brown:1987rra}
F.~R. Brown and T.~J. Woch, ``{Overrelaxed Heat Bath and Metropolis Algorithms
  for Accelerating Pure Gauge Monte Carlo Calculations},''
  \href{http://dx.doi.org/10.1103/PhysRevLett.58.2394}{{\em Phys. Rev. Lett.}
  {\bf 58} (1987)  2394}.

\bibitem{Gupta:1984gq}
R.~Gupta, G.~Guralnik, A.~Patel, T.~Warnock, and C.~Zemach, ``{Monte Carlo
  Renormalization Group for SU(3) Lattice Gauge Theory},''
  \href{http://dx.doi.org/10.1103/PhysRevLett.53.1721}{{\em Phys. Rev. Lett.}
  {\bf 53} (1984)  1721}.

\bibitem{Creutz:1987xi}
M.~Creutz, ``{Overrelaxation and Monte Carlo Simulation},''
  \href{http://dx.doi.org/10.1103/PhysRevD.36.515}{{\em Phys. Rev. D} {\bf 36}
  (1987)  515}.

\bibitem{Francis:2015lha}
A.~Francis, O.~Kaczmarek, M.~Laine, T.~Neuhaus, and H.~Ohno, ``{Critical point
  and scale setting in SU(3) plasma: An update},''
  \href{http://dx.doi.org/10.1103/PhysRevD.91.096002}{{\em Phys. Rev. D} {\bf
  91} (2015) no.~9, 096002}, \href{http://arxiv.org/abs/1503.05652}{{\tt
  arXiv:1503.05652 [hep-lat]}}.

\bibitem{Boyd:1996bx}
G.~Boyd, J.~Engels, F.~Karsch, E.~Laermann, C.~Legeland, M.~Lutgemeier, and
  B.~Petersson, ``{Thermodynamics of SU(3) lattice gauge theory},''
  \href{http://dx.doi.org/10.1016/0550-3213(96)00170-8}{{\em Nucl. Phys. B}
  {\bf 469} (1996)  419--444}, \href{http://arxiv.org/abs/hep-lat/9602007}{{\tt
  arXiv:hep-lat/9602007}}.

\end{thebibliography}

\end{document}